\documentclass[11pt, a4paper]{article}

\usepackage[a4paper, total={16cm, 24cm}]{geometry}
\usepackage{amsmath}
\usepackage{graphicx}
\usepackage{amsthm}
\usepackage{color}
\usepackage{footmisc}

\usepackage{multirow}

\newtheorem{rational for conjecture}{Rational for Conjecture}

\usepackage{algorithm}
\usepackage{algorithmic}
\newcommand{\INPUT}{\textbf{Input: }}

\newcommand{\METHOD}{\textbf{Method: }}

\begin{document}

\LARGE
\begin{center}
\textbf{Joint Tracking of Multiple Quantiles Through Conditional Quantiles} \\[10mm]
\end{center}

\large

\begin{center}
Hugo Lewi Hammer\footnote[2]{OsloMet -- Oslo Metropolitan University}\footnote[4]{Corresponding author. Email: \texttt{hugo.hammer@oslomet.no}}, Anis Yazidi\footnotemark[2] and H{\aa}vard Rue\footnote[5]{King Abdullah University of Science \& Technology}
\end{center}

\normalsize

\begin{abstract}
    Estimation of quantiles is one of the most fundamental real-time analysis tasks. Most real-time data streams vary dynamically with time and incremental quantile estimators document state-of-the art performance to track quantiles of such data streams. However, most are not able to make joint estimates of multiple quantiles in a consistent manner, and estimates may violate the monotone property of quantiles. In this paper we propose the general concept of \emph{conditional quantiles} that can extend incremental estimators to jointly track multiple quantiles. We apply the concept to propose two new estimators. Extensive experimental results, on both synthetic and real-life data, show that the new estimators clearly outperform legacy state-of-the-art joint quantile tracking algorithm and achieve faster adaptivity in dynamically varying data streams.
\end{abstract}


\section{Introduction}

The volumes of automatically generated data are constantly increasing \cite{ramirez2017survey} with more urgent  demand for being analyzed in real-time \cite{kejariwal2015real}. Conventional statistical and data mining techniques are constructed for offline situations and are not applicable for such real-time analysis \cite{krempl2014open}. Thus a wide range of streaming algorithms are continuously being developed addressing a range real-time tasks such as clustering, filtering, cardinality estimation, estimation of moments or quantiles, predictions and anomaly detection \cite{kejariwal2015real}.

Given a stream of data, probably the first and most arguably foundational problem is to describe the data distribution. Quantiles are useful to describe the distribution in a flexible and nonparametric way \cite{luo2016quantiles}. Estimation of quantiles of data streams has been considered for a wide range of applications like portfolio risk measurement in the stock market \cite{gilli2006application,abbasi2013bootstrap}, fraud detection \cite{zhang2008detecting}, signal processing and filtering \cite{stahl2000quantile}, climate change monitoring \cite{zhang2011indices}, SLA violation monitoring \cite{sommers2007accurate,sommers2010multiobjective}, network  monitoring \cite{choi2007quantile,liu2018accurate}, Monte Carlo simulation \cite{wang2016quantiles}, structural health monitoring \cite{gregory2018quantile} and non-parametric statistical testing \cite{lall2015data}. Motivated by the importance and the wide range of applications of streaming quantile estimation, in this paper, we will investigate advancing the state-of-the-art when it comes to simultaneous quantile estimation. 


Suppose that we are interested in estimating the quantile related to some probability $q$. The natural approach is to use the $q$ quantile of the sample distribution. Unfortunately, this conventional approach has clear disadvantages for data streams as computation time and memory requirement are linear to the number of samples received so far from the data stream. Such methods thus are infeasible for large data streams.

Several algorithms have been proposed to deal with those challenges. Most of the proposed methods fall under the category of what can be called histogram or batch based methods. The methods are based on efficiently maintaining a histogram estimate of the data stream distribution such that only a small storage footprint is required. Another ally of methods are the so-called incremental update methods. The methods are based on performing small updates of the quantile estimate every time a new sample is received from the data stream. Generally, the current estimate is a convex combination of the estimate at the previous time step and a quantity depending on the current observation. A thorough review of state-of-the-art streaming quantile estimation methods is given in the related work section (Section \ref{sec:relwork}).

In data stream applications, a common situation is that the distribution of the data stream varies with time. Such system or environment is referred to as a dynamical system in the literature. Given a dynamical system, the problem most commonly addressed is to dynamically update estimates of quantiles of all data received from the data stream so far. Histogram based methods are well suited to address this problem. A less studied, yet important, problem is to estimate quantiles of the current distribution of the data stream typically referred to as quantile tracking. Incremental methods can document state-of-the-performance for the quantile tracking problem \cite{yazidi2017multiplicative,hammer2018a}, while histogram methods are not well suited for efficient quantile tracking \cite{cao2009incremental}. 

To address the tracking problem, several incremental quantile estimators have been suggested \cite{Chambers2006,cao2009incremental,cao2010tracking,yazidi2017multiplicative,ma2013frugal,tiwari2018technique,liu2018accurate}. The intuitions behind the estimators are simple. If the received sample has a value below some threshold, e.g. the current quantile estimate, the estimate is decreased.  Alternatively, whenever the received sample has a value above the same threshold, the estimate is increased. Even though the estimators document state-of-the-art tracking performance \cite{yazidi2017multiplicative}, neither of them use the values of the received samples directly to update the estimate, but only whether the value of the samples are above or below some varying threshold. Intuitively, this seems like a loss of information received from the data stream. Recently, Hammer et al. \cite{hammer2018a,hammer2018novel} presented an incremental estimator that used the values of the received samples directly which makes it distinct from \textit{all} incremental estimators previously presented in the literature. The estimator is in fact a generalized exponentially weighted average of previous observations received from the data stream and documents state-of-the-art performance \cite{hammer2018a,hammer2018novel}. 

The incremental estimators above are constructed to track a single quantile, but from a practical point of view it is often more important to jointly track multiple quantiles, e.g. to be able to approximate the current data stream distribution. Of course one could run multiple incremental estimators in parallel, but we will then loose control over the joint properties of the estimates. Even the monotone property of quantiles\footnote{E.g. that the 70\% quantile estimate must have a value above the 60\% quantile estimate} most likely will be violated. Surprisingly little research has been devoted to develop incremental quantile estimators that are able to make joint estimates of multiple quantiles for dynamically varying data streams. To the best of our knowledge, the only methods in the literature are due to of Cao et al. \cite{cao2009incremental} and Hammer et al. \cite{hammer2017incremental,hammer2017estimation,hammer2018tracking}. The method by Cao et al. is based on first running an incremental update of each quantile estimate and secondly computing a monotonically increasing approximation of the cumulative distribution of the data stream distribution using a form of monotonically increasing linear interpolation. Finally, the quantile estimates are computed from the approximate cumulative distribution. A disadvantage of the method is that the ``monotonization'' approach is quite ad-hoc and cannot give any guarantee that the resulting quantile estimates converge to the true quantiles. The methods by Hammer et al. \cite{hammer2017incremental,hammer2017estimation,hammer2018tracking} are based on adjusting the update step size in each iteration to ensure that the monotone property of quantiles is satisfied. A disadvantage with these methods is that the required step sizes to satisfy the monotone property of quantiles can be too small to efficiently track the changes of the data stream distribution. This will be further demonstrated in the paper. 

The two algorithms presented in this paper address the shortcomings of the current state-of-the-art multiple quantile tracking estimators described above. We show that the suggested estimators converge to the true quantiles and that the estimators can efficiently make joint estimates of multiple quantiles even when the properties of the data stream distribution change rapidly over time. Compared to most streaming quantile estimators in the literature, the suggested algorithms are extremely computationally and memory efficient. In fact, our algorithms require only storing a single value per quantile estimate.

The two algorithms extend the algorithms in \cite{yazidi2017multiplicative} and \cite{hammer2018a} by applying a subtle idea of conditional quantiles. The idea is general and can also be used to obtain joint estimates based on other incremental estimators. 

The remainder of this paper is organized as follows. Section \ref{sec:relwork} is dedicated to reviewing the state-of-the-art. 
In Section \ref{sec:violat}, we motivate the quantile estimators developed
in this paper by first showing how the state-of-the-art incremental estimators in general fail to satisfy the monotone property when estimating multiple quantiles. In Section \ref{sec:joint} we present the general concept of how to obtain joint estimates based on an incremental estimator.  In Sections \ref{sec:ShiftQ} and \ref{sec:CondQ}, we apply the concept to obtain two new algorithms for joint quantile estimation. In Section \ref{sec:exp}, we present comprehensive experimental results that catalogue the properties of our estimators. Section \ref{sec:closrem} concludes the article.

\section{Related Work}
\label{sec:relwork}

The most representative work for this type of ``streaming'' quantile estimator is due to the seminal work of Munro and Paterson \cite{Munro1980}. In \cite{Munro1980}, Munro and Paterson described a $p$-pass algorithm for selection using $O (n^{1/(2p)})$ space for any $p \ge 2$. Cormode and Muthukrishnan \cite{Cormode2005} proposed a more space-efficient data structure, called the Count-Min sketch, which is inspired by Bloom filters, where one estimates the quantiles of a stream as the quantiles of a random sample of the input. The key idea is to maintain a random sample of an appropriate size to estimate the quantile, where the premise is to select a subset of elements whose quantile approximates the true quantile. From this perspective, the latter body of research requires a certain amount of memory that increases as the required accuracy of the estimator increases \cite{Weide1978}. Examples of these works are \cite{Arasu2004, Weide1978, Munro1980, Greenwald2001, guha2009stream, liang2018continuously, liang2019online}. 

In \cite{naumov2007exponentially}, the authors propose a memory efficient method for simultaneous estimation of several quantiles using interpolation methods and a grid structure where each internal grid point is updated upon receiving an observation. The application of this approach is limited for stationary data. The approximation of the quantiles relies on using linear and parabolic interpolations, while the tails of the distribution are approximated using exponential curves. It is worth mentioning that the latter algorithm is based on the $P^2$ algorithm \cite{jain1985p}. In \cite{jain1985p}, Jain et al. resort to five markers so that to track the quantile, where the markers correspond to different quantiles and the min and max of the observations. Their concept is similar to the notion of histograms, where each marker has  two measurements, its height and its position. By definition, each marker has some ideal position, where some adjustments are made to keep it in its ideal position by counting the number of samples exceeding the marker. In simple terms, for example, if the marker corresponds to the $80\%$ quantile, its ideal position will be around the point corresponding to $80\%$ of the data points below the marker. However, such approach does not handle the case of non-stationary quantile estimation as the position of the markers will be affected by stale data points. Then based on the position of the markers, quantiles are computed by supposing that the curve passing through three adjacent markers is parabolic and by using a piecewise parabolic prediction function.

In many network monitoring applications, quantiles are key indicators
for monitoring the performance of the system. For instance, system administrators are interested in monitoring the $95\%$ quantile  of the response time of a web-server so that to hold it under a certain threshold. Quantile tracking is also useful for detecting abnormal events and in intrusion detection systems in general. However, the immense traffic volume of high speed networks impose some computational challenges: little storage and the fact that the computation needs to be ``one pass'' on the data.
It is worth mentioning that the seminal paper of Robbins and Monro \cite{robbins1951stochastic} which established the field of research called  ``stochastic approximation'' \cite{kushner2003stochastic} have included an incremental quantile estimator as a proof of concept of the vast applications of the theory of stochastic approximation.
An extension of the latter quantile estimator which first appeared as example in \cite{robbins1951stochastic} was further developed in \cite{joseph2004efficient} in order to handle the case of ``extreme quantiles''. Moreover, the estimator provided by Tierney \cite{Tierney1983} falls under the same umbrella of the example given in \cite{robbins1951stochastic}, and thus can be seen as an extension of it.

As Arandjelovic remarks \cite{arandjelovic2015two}, most quantile estimation algorithms are not single-pass algorithms and thus are not applicable for streaming data. On the other hand, the  single pass algorithms are concerned with the exact computation of the quantile and thus require a storage space of the order of the size of the data which is clearly an unfeasible condition in the context of big data stream. Thus, we submit that all work on quantile estimation using more than one pass, or storage of the same order of the size of the observations seen so far, is not relevant in the context of this paper.

Given dynamically varying data stream, two main problems are considered namely to i) dynamically update estimates of quantiles of all data received from the stream so far or ii) estimate quantiles of the current distribution of the data stream (tracking). To address problem i), histogram based methods form an important class of memory efficient methods. A representative work in this perspective is due to Schmeiser and Deutsch \cite{schmeiser1977quantile}. In fact, Schmeiser and Deutsch proposed to use equidistant bins where the boundaries are adjusted online. Arandjelovic et al. \cite{arandjelovic2015two} use a different idea than equidistant bins by attempting to maintain bins in a manner that maximizes the entropy of the corresponding estimate of the historical data distribution. Thus, the bin boundaries are adjusted in an online manner.
Nevertheless, histogram based methods have problems addressing problem ii) of tracking quantiles of the current data stream distribution \cite{cao2009incremental}.

To address the dynamic tracking problem ii) incremental algorithms represent an important class of methods. However, the research on incremental methods is quite sparse. As described in the introduction the methods are based in making small updates of the quantile estimates every time a new sample is received. In \cite{Chen2000, cao2010tracking, Chambers2006}, the authors proposed modifications of the stochastic approximation algorithm \cite{Tierney1983}. While Tierney \cite{Tierney1983} uses a sample mean update from previous quantile estimates, several studies \cite{Chen2000, cao2010tracking, cao2009incremental, Chambers2006} propose an exponential decay when taking into account the old estimates. This modification is particularly helpful to track quantiles of non-stationary data stream distributions. Indeed, a ``weighted'' update scheme is applied to incrementally build local approximations of the distribution function in the neighborhood of the quantiles. More recent incremental quantile estimation approaches are the Frugal algorithm by Ma et al. \cite{ma2013frugal}, the DUMIQE algorithm by Yazidi and Hammer \cite{yazidi2017multiplicative}, and the DQTRE and DQTRSE algorithms by Tiwari and Pandey \cite{tiwari2018technique}. A nice property of the DUMIQE and the estimators suggested in this paper is that the update size is automatically adjusted in accordance to the scale/range of the data. This makes the estimators robust to substantial changes in the data stream. The DQTRE and DQTRSE aim to achieve the same aim by estimating the range of the data using peak and valley detectors. However, a disadvantage of these algorithms is that several tuning parameters are required to estimate the range of the data making the algorithms challenging to tune. Furthermore, the incremental algorithms above are constructed to track a single quantile. However, in many practical applications, it is required to track multiple quantiles, e.g. to get an overall non-parametric approximation of the current data stream distribution. As pointed out in the introduction the research on joint tracking multiple quantiles is sparse.

\section{Monotone Property Violations for Incremental Quantile Estimators}
\label{sec:violat}

Let $X_n$ denote a stochastic variable representing possible outcomes from a data stream at time $n$ and let $x_n$ denote a random sample (realization). Further let $f_n(x)$ represent the distribution of $X_n$ and $Q_{n}(q)$ the quantile associated with probability $q$, i.e $P(X_n \leq Q_n(q)) = q$.

The paper will focus on joint tracking of quantiles for $K$ different probabilities $q_1, q_2, \ldots, q_K$. A straight-forward approach would be to run $K$ quantile tracking algorithms in isolation, but in this case, the joint properties of the quantiles will not be taken into account and even the monotone property of quantiles may get violated. We illustrate this using the deterministic based multiplicative incremental quantile estimator (DUMIQE) approach from \cite{yazidi2017multiplicative} as an example. Please see \cite{cao2009incremental} for an illustration for another algorithm. When a new $x_n$ is received from the data stream, the DUMIQE updates the estimates as follows 
\begin{align}
\label{eq:1}
  \begin{split}
   \widehat{Q}_{n+1}(q_k) &\leftarrow (1 + \lambda q_k) \widehat{Q}_{n}(q_k)  \hspace{15mm}\text{ if } \widehat{Q}_{n}(q_k) < x_n \\
   \widehat{Q}_{n+1}(q_k) &\leftarrow (1 - \lambda (1-q_k)) \widehat{Q}_{n}(q_k)  \,\hspace{5mm}\text{ if } \widehat{Q}_{n}(q_k) \geq x_n
  \end{split}
\end{align}
for $k = 1,2,\ldots,K$. Please note that the DUMIQE algorithm assumes that $\widehat{Q}_{n}(q_k) > 0\, \forall\, k,n$ which is not useful if the true quantiles are negative for some $n$. To be able to efficiently track any quantile, \cite{yazidi2017multiplicative} suggested two simple solutions. The first is based on tracking a phantom variable $D_n = X_n + \Delta_n$ where $\Delta_n$ is iteratively updated such that $D_n > Q_{\text{min}} > 0$. The second approach is based on combining the DUMIQE above for positive quantiles and a modified version for negative quantiles. Please see \cite{yazidi2017multiplicative} for further details.

Assume that the monotone property is satisfied and that the sample $x_n$ admits a value between $\widehat{Q}_{n}(q_k)$ and $\widehat{Q}_{n}(q_{k+1})$, i.e
\begin{align}
\label{eq:2}
\widehat{Q}_{n}(q_1) \leq \cdots \leq \widehat{Q}_{n}(q_k) < x_n < \widehat{Q}_{n}(q_{k+1}) \leq \cdots \leq \widehat{Q}_{n}(q_K)
\end{align}
Then according to Equation \eqref{eq:1} the estimates are updated as follows
\begin{align}
\label{eq:3}
  \begin{split}
  \widehat{Q}_{n+1}(q_j) &\leftarrow (1 + \lambda q_j) \widehat{Q}_{n}(q_j) \hspace{20mm}\text{ for } j = 1,2, \ldots, k\\
  \widehat{Q}_{n+1}(q_{j}) &\leftarrow (1 - \lambda (1-q_{j})) \widehat{Q}_{n}(q_{j}) \,\hspace{10mm}\text{ for } j = k+1, \ldots, K
  \end{split}
\end{align}
which means that the estimates are increased for the quantiles with an estimate below $x_n$ and decreased for the estimates above $x_n$. Consequently, the monotone property may get violated. 

\section{Joint Tracking of Multiple Quantiles}
\label{sec:joint}

In Sections \ref{sec:ShiftQ} and \ref{sec:CondQ} we will present two new algorithms for joint tracking of multiple quantiles based on extending the DUMIQE and QEWA algorithms \cite{hammer2018a}, respectively.  

Both algorithms are based on the same concept: first track a central quantile of the distribution and typically the median. Next track other quantiles \textit{relative} to the central quantile by taking advantage of conditional distributions. E.g. to track the 25\% quantile, simply track the median of observations below the estimates of the median. However, some more sophisticated modifications of this concept will be conducted before obtaining the final algorithms. 

Please note that the concept is general and can be applied to also extend other incremental tracking algorithms then the DUMIQE and QEWA considered in this paper.

The algorithms will take advantage of the following properties. 

\textbf{Property 1:} Define $Y$ as a shifted variable of $X$, $X = Y + \delta$ and let $Q_Y(q)$ denote the $q$ quantile of $Y$, then $Q_X(q) = Q_Y(q) + \delta$. This means that if $Y$ is a shifted variable of $X$, a similar shift is observed in the quantiles. This follows from
\begin{align*}
  P(X < Q_X(q)) = q = P(Y < Q_Y(q)) = P(Y + \delta < Q_Y(q) + \delta) = P(X < Q_Y(q) + \delta)
\end{align*}

\textbf{Property 2:} Let $q_1 < q_2$. Consider the conditional probability
\begin{align}
  \label{eq:4}
  P(X < Q_X(q_1)\, |\, X < Q_X(q_2)) = \frac{P(X < Q_X(q_1))}{P(X < Q_X(q_2))} = \frac{q_1}{q_2}
\end{align}
which means that the $q_1$ quantile of $X$, $Q_X(q_1)$, is equal to the $q_1/q_2$ quantile of the conditional variable $X\, |\, X<Q_X(q_2)$ with truncated distribution $f(x\, |\, x<Q_X(q_2))$. Applying \eqref{eq:4} on a shifted variable $Y = X - Q_X(q_2)$ gives
\begin{align}
  \label{eq:12}
  P(Y < Q_Y(q_1) \,|\, Y < 0) = \frac{q_1}{q_2}
\end{align}
using that $Q_Y(q_2) = 0$ (Property 1). This means that the $q_1$ quantile of $Y$ is equal to the $q_1/q_2$ quantile of the conditional variable $Y\, |\, Y<0$.

Conditioning in the opposite direction 
\begin{align}
  \label{eq:5}
  P(X < Q_X(q_2)\, |\, X > Q_X(q_1)) = \frac{P(X < Q_X(q_2) \cap X > Q_X(q_1))}{P(X > Q_X(q_1))} = \frac{q_2 - q_1}{1 - q_1}
\end{align}
which means that the $q_2$ quantile of $X$ is equal to the $(q_2 - q_1)/(1 - q_1)$ quantile of the conditional variable $X\, |\, X > Q(q_1)$ with truncated distribution $f(x\, |\, x > Q(q_1))$. Applying \eqref{eq:5} on a shifted variable $Y = X - Q_X(q_1)$ gives
\begin{align}
  \label{eq:13}
  P(Y < Q_Y(q_2) \,|\, Y > 0) = \frac{q_2 - q_1}{1 - q_1}
\end{align}
using that $Q_Y(q_1) = 0$ (Property 1). This means that the $q_2$ quantile of $Y$ is equal to the $(q_2 - q_1)/(1 - q_1)$ quantile of the conditional variable $Y\, |\, Y > 0$.

\textbf{Property 3:} Finally we will take advantage of the multiplicative nature of DUMIQE (Equation \eqref{eq:1}), which means that if $\widehat{Q}_{0}(q) > $ all following estimates will be strictly positive. 

Due to the multiplicative nature of the DUMIQE algorithm (Property 3), we can jointly track multiple quantiles without using Property 2. This simplifies the algorithm. We also tested an algorithm based on DUMIQE using both Properties 1 and 2 resulting in a more complicated algorithm without any improved results. The QEWA algorithm is not multiplicative and the extension will use both Properties 1 and 2.

\section{Extension of the DUMIQE Algorithm}
\label{sec:ShiftQ}

We start by presenting the algorithm for $K = 2$ before extending to $K > 2$. 

\subsection{Tracking of Two Quantiles}
\label{sec:dumiqetwo}

Assume that $q_1 < q_2$ and $Q_n(q_1) > 0, n = 1,2,\ldots$ \footnote{If we are not sure that this is satisfied, a transformation as described in Section \ref{sec:violat} can be used.}. The algorithm will take advantage of a shifted variable $Y_{n,1} = X_n - \widehat{Q}_{n+1}(q_1)$. Let $\widehat{Q}_{Y,n}(q_2)$ denote an estimate of the $q_2$ quantile of $Y_{n,1}$. The monotone property $\widehat{Q}_{n}(q_1) < \widehat{Q}_{n}(q_2)$ implies that $\widehat{Q}_{Y,n}(q_1) > 0$ (Property 1) and we thus require $\widehat{Q}_{Y,n}(q_2) > 0$. The algorithm is initiated with $\widehat{Q}_{0}(q_1) > 0$ and $\widehat{Q}_{Y,0}(q_2) > 0$ and consists of the following updates:
\begin{itemize}
\item[1.1] Update $\widehat{Q}_{n}(q_1)$ using the DUMIQE update rule in Equation \eqref{eq:1}
\begin{align*}
   \widehat{Q}_{n+1}(q_1) &\leftarrow (1 + \lambda q_1) \widehat{Q}_{n}(q_1)  \hspace{15mm}\text{ if } \widehat{Q}_{n}(q_1) < x_n \\
   \widehat{Q}_{n+1}(q_1) &\leftarrow (1 - \lambda (1-q_1)) \widehat{Q}_{n}(q_1)  \,\hspace{5mm}\text{ if } \widehat{Q}_{n}(q_1) \geq x_n
\end{align*}
\item[1.2] Compute the shifted observation $y_{n,1} \leftarrow x_n - \widehat{Q}_{n+1}(q_1)$.
\item[1.3] Track the $q_2$ quantile of the shifted variable
\begin{align*}
   \widehat{Q}_{Y,n+1}(q_2) &\leftarrow (1 + \gamma q_2) \widehat{Q}_{Y,n}(q_2)  \hspace{15mm}\text{ if } \widehat{Q}_{Y,n}(q_2) < y_{n,1} \\
   \widehat{Q}_{Y,n+1}(q_2) &\leftarrow (1 - \gamma (1-q_2)) \widehat{Q}_{Y,n}(q_2)  \,\hspace{5mm}\text{ if } \widehat{Q}_{Y,n}(q_2) \geq y_{n,1}
\end{align*}
\item[1.4] Finally get an estimate for $Q_{n+1}(q_2)$ by shifting back (Property 1).
  \begin{align}
    \label{eq:10}
    \widehat{Q}_{n+1}(q_2) \leftarrow \widehat{Q}_{Y,n+1}(q_2) + \widehat{Q}_{n+1}(q_1) 
  \end{align} 
\end{itemize}
It is straight forward to see that the monotone property is satisfied. Due to the multiplicative update form of the DUMIQE, $\widehat{Q}_{Y,n}(q_2)$ is positive for every $n$. Thus from Equation \eqref{eq:10} it follows that $\widehat{Q}_{n}(q_1) < \widehat{Q}_{n}(q_2)$ for every $n$.

The quantiles can further be updating in the opposite direction, i.e. by first tracking the $Q_n(q_2)$ quantile of the data stream and then $Q_n(q_1)$ relative to the $Q_n(q_2)$. We assume that $Q_n(q_2) > 0, n = 1,2,\ldots$ and again introduce a shifted variable, but in addition change sign $Y_{n,2} = \widehat{Q}_{n+1}(q_2) - X_n$. Let $\widehat{Q}_{Y,n}(q_1)$ denote an estimate of the $q_1$ quantile of $Y_{n,2}$. The algorithm is initiated with  $\widehat{Q}_{0}(q_2) > 0$ and $\widehat{Q}_{Y,0}(q_1) > 0$ and consists of the following updates:
\begin{itemize}
\item[2.1] Update $\widehat{Q}_{n}(q_2)$ using the DUMIQE update rule in Equation \eqref{eq:1}
\begin{align*}
   \widehat{Q}_{n+1}(q_2) &\leftarrow (1 + \lambda q_2) \widehat{Q}_{n}(q_2)  \hspace{15mm}\text{ if } \widehat{Q}_{n}(q_2) < x_n \\
   \widehat{Q}_{n+1}(q_2) &\leftarrow (1 - \lambda (1-q_2)) \widehat{Q}_{n}(q_2)  \,\hspace{5mm}\text{ if } \widehat{Q}_{n}(q_2) \geq x_n
\end{align*}
\item[2.2] Compute the shifted observation $y_{n,2} \leftarrow \widehat{Q}_{n+1}(q_2) - x_n$. 
\item[2.3] Track the $q_1$ quantile of the shifted variable using DUMIQE
\begin{align*}
   \widehat{Q}_{Y,n+1}(q_1) &\leftarrow (1 + \gamma q_1) \widehat{Q}_{Y,n}(q_1)  \hspace{15mm}\text{ if } \widehat{Q}_{Y,n}(q_1) < y_{n,2} \\
   \widehat{Q}_{Y,n+1}(q_1) &\leftarrow (1 - \gamma (1-q_1)) \widehat{Q}_{Y,n}(q_1)  \,\hspace{5mm}\text{ if } \widehat{Q}_{Y,n}(q_1) \geq y_{n,2}
\end{align*}
\item[2.4] Finally get an estimate for $Q_{n+1}(q_1)$ by shifting back (Property 1).
  \begin{align}
    \label{eq:11}
    \widehat{Q}_{n+1}(q_1) \leftarrow \widehat{Q}_{n+1}(q_2) - \widehat{Q}_{Y,n+1}(q_1) 
  \end{align} 
\end{itemize}
Again since $\widehat{Q}_{Y,n+1}(q_2)$ is positive it follows from \eqref{eq:11} that $\widehat{Q}_{n}(q_1) < \widehat{Q}_{n}(q_2)$.

\subsection{Tracking of Multiple Quantiles}
\label{sec:shifttwo}

The algorithm for a general $K$ is shown in Algorithm \ref{alg:2}. In step 2, $q_c$ refers to a central probability of the data stream distribution and typically $q_c = 0.5$ such that $\widehat{Q}_{n}(q_c)$ is an estimate of the median. The function $\text{DUMIQE}(\widehat{Q}_{n}(q_c), x_n, q_c, \lambda)$ refers to one update with the DUMIQE algorithm with $\widehat{Q}_{n}(q_c)$ referring to the estimate of $Q_n(q_c)$, $x_n$ the received data stream observation and $\lambda$ the value of the tuning parameter.

In steps 3 to 7, the procedure 2.1 to 2.4 in the previous section is repeatedly performed. First $\widehat{Q}_{n+1}(q_{c-1})$ is updated relative to $\widehat{Q}_{n+1}(q_c)$, then $\widehat{Q}_{n+1}(q_{c-2})$ relative to $\widehat{Q}_{n+1}(q_{c-1})$ and so on. In steps 8 to 12, the procedure 1.1 to 1.4 in is repeatedly performed.

\begin{algorithm}
\caption{Joint quantiles tracking algorithm based on the an extension of DUMIQE.}\label{alg:2}
\INPUT\\
$x_1, x_2, x_3, \ldots$ // Data stream\\
$\lambda$, $\gamma$, $K$, $c$ \\
$0 < \widehat{Q}_{0}(q_1) < \cdots < \widehat{Q}_{0}(q_K)$\\
$0 < \widehat{Q}_{Y,0}(q_1) < \cdots < \widehat{Q}_{Y,0}(q_K)$\\
\METHOD
\begin{algorithmic}[1]
  \FOR {$n \in 1,2,\ldots$}
    \STATE $\widehat{Q}_{n+1}(q_c) \leftarrow \text{DUMIQE}(\widehat{Q}_{n}(q_c), x_n, q_c, \lambda)$
    \FOR {$k \in c-1,\ldots,1$} 
        \STATE $y_{n,k+1} \leftarrow \widehat{Q}_{n+1}(q_{k+1}) - x_n$
        \STATE $\widehat{Q}_{Y,n+1}(q_k) \leftarrow \text{DUMIQE}(\widehat{Q}_{Y,n}(q_k), y_{n,k+1}, q_k, \gamma)$
        \STATE $\widehat{Q}_{n+1}(q_k) \leftarrow  \widehat{Q}_{n+1}(q_{k+1})  - \widehat{Q}_{Y,n+1}(q_k)$
    \ENDFOR
    \FOR {$k \in c+1,\ldots,K$} 
        \STATE $y_{n,k+1} \leftarrow x_n - \widehat{Q}_{n+1}(q_{k-1})$
        \STATE $\widehat{Q}_{Y,n+1}(q_k) \leftarrow \text{DUMIQE}(\widehat{Q}_{Y,n}(q_k), y_{n,k-1}, q_k, \gamma)$
        \STATE $\widehat{Q}_{n+1}(q_k) \leftarrow \widehat{Q}_{Y,n+1}(q_k) + \widehat{Q}_{n+1}(q_{k-1})$
    \ENDFOR
  \ENDFOR
\end{algorithmic}
\end{algorithm}

\section{Extension of the QEWA Algorithm}
\label{sec:CondQ}

The QEWA algorithm consists of the following updates. 
\begin{align}
  &\bullet\hspace{4mm} \label{eq:8}\widehat{Q}_{n+1}(q) \leftarrow (1 - \widehat{b}_n) \widehat{Q}_{n}(q) + \widehat{b}_n x_n \\[2mm]
  &\notag \bullet\hspace{4mm} \text{If } x_n > \widehat{Q_{n}}(q)\\
  &\hspace{8mm}\text{-}\hspace{4mm} \label{eq:6}\widehat{\mu}_{n+1}^+ \leftarrow \widehat{Q}_{n+1}(q) - \widehat{Q}_{n}(q) + (1-\rho) \widehat{\mu}_{n}^+ + \rho x_n\\
  &\hspace{8mm}\text{-}\hspace{4mm} \label{eq:6b}\widehat{\mu}_{n+1}^- \leftarrow \widehat{Q}_{n+1}(q) - \widehat{Q}_{n}(q) + \widehat{\mu}_{n}^- \\
  &\notag \bullet\hspace{4mm}\text{Else}\\
  &\hspace{8mm}\text{-}\hspace{4mm} \label{eq:7b}\widehat{\mu}_{n+1}^+ \leftarrow \widehat{Q}_{n+1}(q) - \widehat{Q}_{n}(q) + \widehat{\mu}_{n}^+ \\
  &\hspace{8mm}\text{-}\hspace{4mm} \label{eq:7}\widehat{\mu}_{n+1}^- \leftarrow \widehat{Q}_{n+1}(q) - \widehat{Q}_{n}(q) + (1-\rho) \widehat{\mu}_{n}^- + \rho x_n \\
  &\bullet\hspace{4mm} \widehat{a}_{n+1} \leftarrow \frac{q}{\widehat{\mu}_{n+1}^+ - \widehat{Q}_{n+1}(q)} \left/ \left( \frac{q}{\widehat{\mu}_{n+1}^+ - \widehat{Q}_{n+1}(q)} + \frac{1-q}{\widehat{Q}_{n+1}(q) - \widehat{\mu}_{n+1}^-} \right ) \right.\\
  &\bullet\hspace{4mm} \label{eq:9} \widehat{b}_{n+1} \leftarrow \lambda\left(\widehat{a}_{n+1} + I\left(x_n \leq \widehat{Q}_{n+1}(q)\right)(1-2\widehat{a}_{n+1})\right)
\end{align}
where $\widehat{\mu}_{n+1}^+$ and $\widehat{\mu}_{n+1}^-$ represent estimates of the conditional expectations $\mu^+ = E(X_n|X_n > \widehat{Q}_{n}(q))$ and $\mu^- = E(X_n|X_n < \widehat{Q}_{n}(q))$, respectively. From Equation \eqref{eq:8} we see that the estimator is in fact a generalized EWA with weights $0 < b_n < 1$. The weights, $b_n$, are computed such that the estimator tracks the quantile $Q_n(q)$ and not the expectation $E(X_n)$ of the data stream distribution\footnote[3]{If constants weights were used, i.e. $b_n = b$, Equation \eqref{eq:8} would track the expectation $E(X_n)$ and not $Q_n(q)$}. For more details, we refer to \cite{hammer2018a}.

We start by presenting the algorithm for $K = 2$ quantiles before extending to a general $K$. 

\subsection{Tracking of Two Quantiles}
\label{sec:condtwo}

The procedures will consist of the same steps as in Section \ref{sec:shifttwo}, except that steps 1.3 and 2.3 will use conditional quantiles (Property 2) and thus are slightly more involved. The two procedures will be presented in the opposite order compared to above. 

We take advantage of a shifted variable $Y_{n,2} = X_n - \widehat{Q}_{n+1}(q_2)$, which is assumed to be negative $\widehat{Q}_{Y,n}(q_1)$ (Property 1) \footnote{The $Y$'s can be defined in different ways, but we find this to be the easiest.}. The algorithm is initiated with $\widehat{Q}_{0}(q_2) > 0$ and $\widehat{Q}_{Y,0}(q_1) < 0$ and consists of the following updates:\footnote{Please note that to be able to track a quantile using the QEWA algorithm, the conditional expectations must also be tracked to obtain the weights $b_n$. To simplify the explanation, we only focus on the quantile tracking for now.}
\begin{itemize}
\item[1.1] Update $\widehat{Q}_{n}(q_2)$ using the QEWA update rules in Equations \eqref{eq:8} to \eqref{eq:9} to obtain an estimate $\widehat{Q}_{n+1}(q_2)$.
\item[1.2] Compute the shifted observation $y_{n,2} \leftarrow x_n - \widehat{Q}_{n+1}(q_2)$. 
\item[1.3]  Track $Q_{Y,n}(q_1)$ by tracking the $q_1/q_2$ quantile of the conditional variable $Y_{n,2}\, |\, Y_{n,2} < 0$ (recall Equation \eqref{eq:12}). Thus if $y_{n,2} < 0$, apply the update rules in Equations \eqref{eq:8} to \eqref{eq:9} with $q = q_1/q_2$ to obtain a quantile estimate $\widehat{Q}_{Y,n+1}(q_1)$.
Further, if $y_{n,2} > 0$, we are outside of the support of the conditional variable and thus do no update of the quantile estimate, i.e. $\widehat{Q}_{Y,n+1}(q_1) \leftarrow \widehat{Q}_{Y,n}(q_1)$.
\item[1.4] Finally shift back (Property 1).
  \begin{align*}
    \widehat{Q}_{n+1}(q_1) \leftarrow \widehat{Q}_{Y,n+1}(q_1) + \widehat{Q}_{n+1}(q_2) 
  \end{align*}
\end{itemize}
We can now prove that the algorithm ensures the monotone property, $\widehat{Q}_{n}(q_1) < \widehat{Q}_{n}(q_2)$, in every iteration. We start proving, by induction, that if $\widehat{Q}_{Y,0}(q_1) < 0$, every $\widehat{Q}_{Y,n}(q_1)$ will also be negative. Assume that $\widehat{Q}_{Y,n}(q_1) < 0$. If $y_{n,2} < 0$, $\widehat{Q}_{Y,n+1}(q_1)$ is computed using Equation \eqref{eq:8}. Since both $\widehat{Q}_{Y,n}(q_1) < 0$ and $y_{n,2} < 0$, consequently, $\widehat{Q}_{Y,n+1}(q_1) < 0$ (convex combination of two negative values). If $y_{n,2} > 0$, then $\widehat{Q}_{Y,n+1}(q_1) \leftarrow \widehat{Q}_{Y,n}(q_1)$, which again implies $\widehat{Q}_{Y,n+1}(q_1) < 0$. From Step 1.4 this again implies that $\widehat{Q}_{n}(q_1) < \widehat{Q}_{n}(q_2)$. 

The same quantiles can now be updated in the opposite direction. Let $Y_{n,1} = X_n - \widehat{Q}_{n+1}(q_1)$ and initiate the algorithm with $\widehat{Q}_{0}(q_1) > 0$ and $\widehat{Q}_{Y,0}(q_2) > 0$. The algorithm consists for the following updates:
\begin{itemize}
\item[2.1] Update $\widehat{Q}_{n}(q_1)$ using the QEWA update rules in Equations \eqref{eq:8} to \eqref{eq:9} to obtain an estimate $\widehat{Q}_{n+1}(q_1)$.
\item[2.2] Compute the shifted observation $y_{n,1} \leftarrow x_n - \widehat{Q}_{n+1}(q_1)$.
\item[2.3]  Track $Q_{Y,n}(q_2)$ by tracking the $(q_2 - q_1)/(1 - q_1)$ quantile of the conditional variable $Y_{n,1}\, |\, Y_{n,1} > 0$ (recall Equation \ref{eq:13}). Thus if $y_{n,1} > 0$, we update using Equations \eqref{eq:8} to \eqref{eq:9} with $q = (q_2 - q_1)/(1 - q_1)$ to obtain a quantile estimate $\widehat{Q}_{Y,n+1}(q_2)$. If $y_{n,1} < 0$, no update: $\widehat{Q}_{Y,n+1}(q_2) \leftarrow \widehat{Q}_{Y,n}(q_2)$.
\item[2.4] Shift back (Property 1).
  \begin{align*}
    \widehat{Q}_{n+1}(q_2) \leftarrow \widehat{Q}_{Y,n+1}(q_2) + \widehat{Q}_{n+1}(q_1) 
  \end{align*} 
\end{itemize}
By the same reasoning as above, also this algorithm ensures the monotone property.

\subsection{Tracking of Multiple Quantiles}
\label{sec:multiple}

The algorithm for a general $K$ is shown in Algorithm \ref{alg:1} where $\widehat{\mu}_{c,n}^-$ and $\widehat{\mu}_{c,n}^+$ refer to estimates of the conditional expectations $E(X_n|X_n < \widehat{Q}_{n}(q_c))$ and $E(X_n|X_n > \widehat{Q}_{n}(q_c))$, respectively. Further $\widehat{\mu}_{Y,k,n}^-$ and $\widehat{\mu}_{Y,k,n}^+$ refer to estimates of the conditional expectations $E(Y_{n,k}|Y_{n,k} < \widehat{Q}_{Y,n}(q_k))$ and $E(Y_{n,k}|Y_{n,k} > \widehat{Q}_{Y,n}(q_k))$, respectively. The function $\text{QEWA}(\widehat{Q}_{n}(q_c), \widehat{\mu}_{c,n}^-, \widehat{\mu}_{c,n}^+, x_n, q_c, \lambda, \rho)$ refers to one update with the QEWA algorithm with $\widehat{Q}_{n}(q_c), \widehat{\mu}_{c,n}^-$ and $\widehat{\mu}_{c,n}^+$ representing estimates of $Q_n(q_c)$, $E(X_n|X_n < \widehat{Q}_{n}(q_c))$ and $E(X_n|X_n > \widehat{Q}_{n}(q_c))$, respectively, $x_n$ the data stream observation and $\lambda$ and $\rho$ the tuning parameter of the QEWA algorithm. In steps 3 to 11 and in steps 12 to 21, the procedures ind 1.1 to 1.4 and 2.1 to 2.4, in the previous section, are repeatedly performed.
\begin{algorithm}
\caption{Joint quantiles tracking algorithm based on the an extension of QEWA.}\label{alg:1}
\INPUT\\
$x_1, x_2, x_3, \ldots$ // Data stream\\
$\lambda$, $\gamma$, $\rho$, $K$, $c$ \\
$\widehat{\mu}_{c,0}^- < \widehat{Q}_{0}(q_c) < \widehat{\mu}_{c,0}^+$\\
$\widehat{\mu}_{Y,k,0}^- < \widehat{Q}_{Y,0}(q_k) < \widehat{\mu}_{Y,k,0}^+ < 0, \,\, k < c$\\
$0 < \widehat{\mu}_{Y,k,0}^- < \widehat{Q}_{Y,0}(q_k) < \widehat{\mu}_{Y,k,0}^+, \,\, k > c$\\
\METHOD
\begin{algorithmic}[1]
  \FOR {$n \in 1,2,\ldots$}
    \STATE $(\widehat{Q}_{n+1}(q_c), \widehat{\mu}_{c,n+1}^-, \widehat{\mu}_{c,n+1}^+) \leftarrow \text{QEWA}(\widehat{Q}_{n}(q_c), \widehat{\mu}_{c,n}^-, \widehat{\mu}_{c,n}^+, x_n, q_c, \lambda, \rho)$
    \FOR {$k \in c-1,\ldots,1$} 
      \IF {$x_n < \widehat{Q}_{n+1}(q_{k+1})$}
        \STATE $y_{n,k+1} \leftarrow x_n - \widehat{Q}_{n+1}(q_{k+1})$
        \STATE $(\widehat{Q}_{Y,n+1}(q_k), \widehat{\mu}_{Y,k,n+1}^-, \widehat{\mu}_{Y,k,n+1}^+) \leftarrow \text{QEWA}(\widehat{Q}_{Y,n}(q_k), \widehat{\mu}_{Y,k,n}^-, \widehat{\mu}_{Y,k,n}^+, y_{n,k+1}, q_{k+1}/q_k, \gamma, \rho)$
      \ELSE
        \STATE $(\widehat{Q}_{Y,n+1}(q_k), \widehat{\mu}_{Y,k,n+1}^-, \widehat{\mu}_{Y,k,n+1}^+) \leftarrow (\widehat{Q}_{Y,n}(q_k), \widehat{\mu}_{Y,k,n}^-, \widehat{\mu}_{Y,k,n}^+)$
      \ENDIF
        \STATE $\widehat{Q}_{n+1}(q_k) \leftarrow \widehat{Q}_{Y,n+1}(q_k) + \widehat{Q}_{n+1}(q_{k+1})$
    \ENDFOR
    \FOR {$k \in c+1,\ldots,K$} 
      \IF {$x_n > \widehat{Q}_{n+1}(q_{k-1})$}
        \STATE $y_{n,k-1} \leftarrow x_n - \widehat{Q}_{n+1}(q_{k-1})$
        \STATE $(\widehat{Q}_{Y,n+1}(q_k), \widehat{\mu}_{Y,k,n+1}^-, \widehat{\mu}_{Y,k,n+1}^+) \leftarrow$
        \STATE $\hspace{32mm}\text{QEWA}(\widehat{Q}_{Y,n}(q_k), \widehat{\mu}_{Y,k,n}^-, \widehat{\mu}_{Y,k,n}^+, y_{n,k-1}, (q_{k} - q_{k-1})/(1 - q_{k-1}), \gamma, \rho)$
      \ELSE
        \STATE $(\widehat{Q}_{Y,n+1}(q_k), \widehat{\mu}_{Y,k,n+1}^-, \widehat{\mu}_{Y,k,n+1}^+) \leftarrow (\widehat{Q}_{Y,n}(q_k), \widehat{\mu}_{Y,k,n}^-, \widehat{\mu}_{Y,k,n}^+)$
      \ENDIF
      \STATE $\widehat{Q}_{n+1}(q_k) \leftarrow \widehat{Q}_{Y,n+1}(q_k) + \widehat{Q}_{n+1}(q_{k-1})$
    \ENDFOR
  \ENDFOR
\end{algorithmic}
\end{algorithm}

We end this section with a few remarks.

\textit{Remark 1:} The estimate $\widehat{Q}_{n}(q_c)$, in Algorithms \ref{alg:2} and \ref{alg:1}, tracks the overall trend of the data stream, while the other quantiles are updated relative to $\widehat{Q}_{n}(q_c)$ and only need to track changes in \textit{shape} of the distribution. Thus for most dynamic data streams it is natural to use a step size $\lambda$ that is on a larger scale than $\gamma$. Our experiments will show that the performance of the algorithm is not sensitive on the value of the $\gamma$ parameter. Any value of $\gamma$ somewhere between $1/10$ and $1/10000$ performed well in all our experiments. The performance of Algorithm \ref{alg:1} is not sensitive on the value of $\rho$ either. Following the recommendation in \cite{hammer2018a} of using $\rho = 0.01 \lambda$ performed well in all our experiments.

\textit{Remark 2:} In \cite{yazidi2017multiplicative} and \cite{hammer2018a} theoretical results are given proving that the DUMIQE and QEWA estimators converge to the true quantiles for static data streams. Although the DUMIQE and QEWA algorithms are designed to track quantiles for dynamic environments, it is an important requirement that the estimators converge to the true quantile for static data streams. These theoretical results again imply that Algorithms \ref{alg:2} and \ref{alg:1} will converge to the true quantiles for static data streams. 

\section{Experiments}
\label{sec:exp}

In this section, we evaluate the performance of Algorithms \ref{alg:2} and \ref{alg:1} for both synthetic and real life data sets. We will denote Algorithms \ref{alg:2} and \ref{alg:1} by ShiftQ and CondQ, respectively (since computation of quantiles are based on SHIFTed and CONDitional stochastic variables). We compare against the alternative multiple quantile tracking algorithm we are aware of in the literature, namely the method of Cao et al. in \cite{cao2009incremental} and the MDUMIQE by Hammer et al. in \cite{hammer2018tracking}. 

\subsection{Synthetic Experiments}

\begin{figure}
  \centering
  \includegraphics[width = \textwidth]{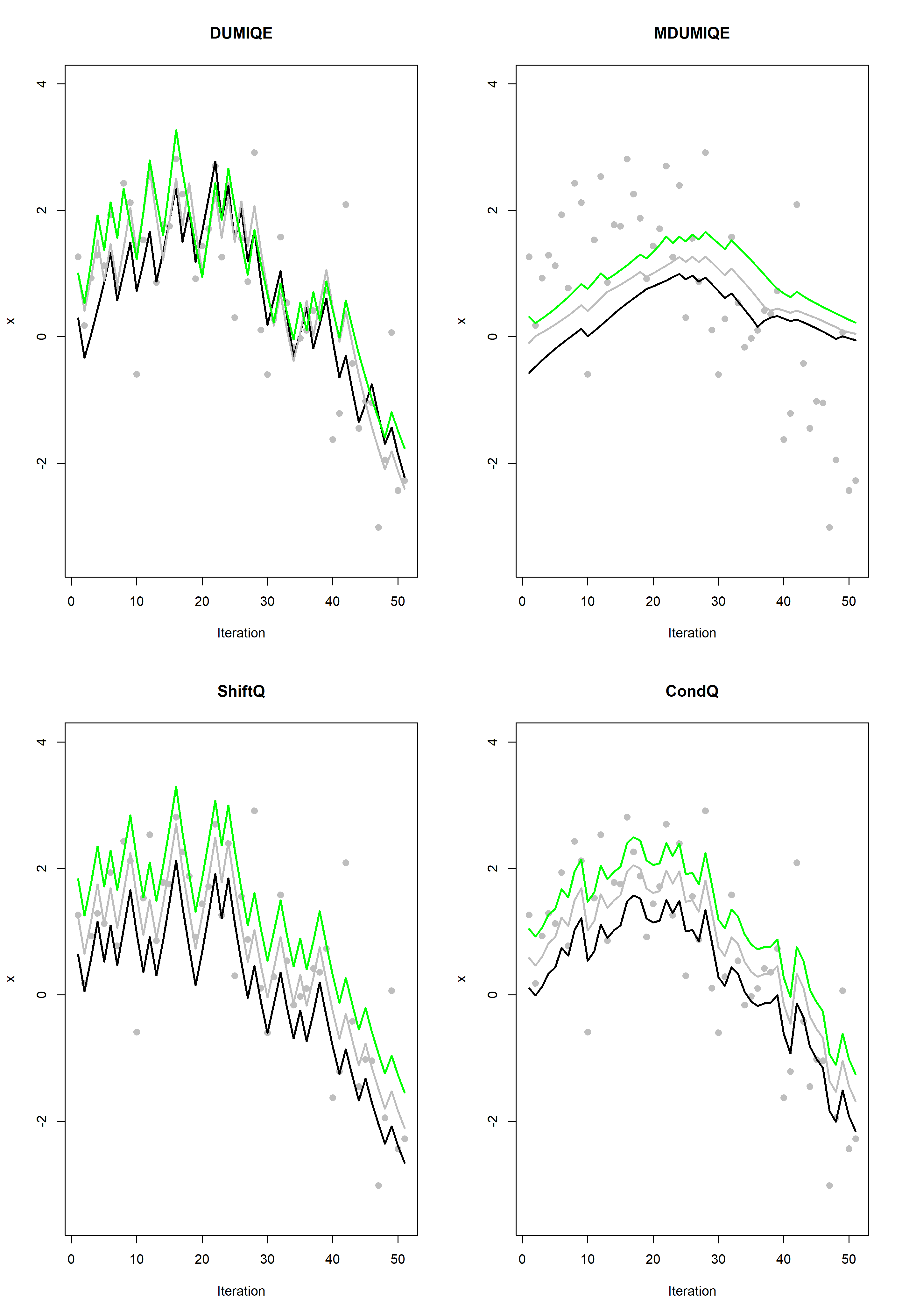}
  \caption{Estimation processes for DUMIQE, MDUMIQE, CondQ and ShiftQ. The gray dots show samples from the data stream distribution while the black, gray and green curves show estimates of the 0.4, 0.5 and the 0.6 quantiles of the data stream, respectively.}
  \label{fig:1}
\end{figure}

Figure \ref{fig:1} shows a small section of the tracking processes for the algorithm DUMIQE, MDUMIQE, CondQ and ShiftQ. The gray dots show the data stream, that is the same in all the four panels. The black, gray and green curves show tracking of the 0.4, 0.5 and the 0.6 quantiles of the data, respectively.
We see that using DUMIQE, the monotone property is violated while the three other algorithms are able to satisfy the monotone property. The idea behind the MDUMIQE is to reduce the step size to avoid monotone property violations. Consequently, as shown on the upper right panel, if the data stream changes rapidly, the MDUMIQE is not able to track efficiently since the adjusted step sizes become too small. We see that both CondQ and ShiftQ are able to efficiently track the dynamics of the data stream. However, CondQ is able to track the dynamics with less noise and thus in total seems the most efficient algorithm.

We now proceed to a more thorough evaluation of the suggested algorithms. The algorithms are designed to perform well for dynamically changing data streams and the experiments will focus on this. We considered four different data cases. For the first case, the data stream distributions were normally distributed and the expectations, $\mu_n$, varied smoothly as follows
\begin{align*}
\mu_n = a \sin \left( \frac{2\pi}{T} n \right), \,\,\, n = 1,2,3, \ldots
\end{align*}
which is a sinus function with period $T$. For the second case, the data stream distributions were also normally distributed, but the expectation switched between values $a$ and $-a$
\begin{align*}
  \mu_n = \left \{
  \begin{array}{ll}
    a & \text{ if } n\,\text{mod}\,T \leq T/2 \\
     -a & \text{ else }
  \end{array}
  \right.
\end{align*}
The standard deviation was set to one. For the two remaining cases, the data stream distributions were $\chi^2$ distributed, one with smooth changes and one with rapid swithces. For the smooth case the number of degrees of freedom, $\nu_n$, varied with time as follows
\begin{align*}
  \nu_n = a \sin \left( \frac{2\pi}{T} n \right) + b, \,\,\, n = 1,2,3, \ldots
\end{align*}
where $b > a$ such that $\nu_n > 0$ for all $n$. For the switch case, the number of degrees of freedom switched between values $a + b$ and $-a + b$
\begin{align*}
  \mu_n = \left \{
  \begin{array}{ll}
    a+b & \text{ if } n\,\text{mod}\,T \leq T/2 \\
    -a+b & \text{ else }
  \end{array}
  \right.
\end{align*}
In the experiments we used $a = 2$ and $b=6$. The $\chi^2$ distribution is quite heavy tailed and both the scale and the shape of the distribution change with time making this a challenging quantile tracking problem.

We conidered the two periods $T=100$ (rapid variation) and $T=1000$ (slow variation). For each data stream, either $K = 3$ or $K = 19$ quantiles were tracked. For the $K = 3$ case, quantiles associated with the probabilities $q_1 = 0.2, q_2 = 0.5$ and $q_3 = 0.8$ quantiles were tracked. For the $K = 19$ case, quantiles associated with the probabilities $q_k = 0.05k, k = 1,\ldots,19$ were tracked.

Tracking error was measured using the average root mean squared error for the different quantiles
\begin{align*}
  \text{RMSE }=  \frac{1}{K} \sum_{k=1}^K  \sqrt{ \frac{1}{N}\sum_{n=1}^N \left(Q_n(q_k) - \widehat{Q}_n(q_k)\right)^2 }
\end{align*}
where $N$ is the total number of samples received from the data stream. In the experiments, we used $N = 10^6$ which efficiently removed any Monte Carlo errors. Following the recommendations in \cite{hammer2018a}, for the QEWA estimator in the CondQ algorithm we used a $\rho/\lambda$ ratio equal to 1/100.

In order to obtain a good overview of the performance of the algorithms, we measured the estimation error for a wide range of values for the step length in the algorithms. However, in a practical situation, the history of the data stream can be used to track optimal values of the tuning parameters\footnote{We are currently working on such procedures}. Thus the focus will be on the performance of the algorithms using optimal step lengths. Complete results, showing estimation errors for every choice of the tuning parameters, are given in Figures \ref{fig:2} to \ref{fig:5} in Appendix \ref{app:synt}.

Tables \ref{tab:5} to \ref{tab:8} show estimation error for the different algorithms using an optimal step length. We see that the CondQ outperforms all the other algorithms for each of the 16 cases (data streams) and mostly with a clear margin. For most of the cases, both CondQ and ShiftQ outperform the algorithm by Cao et al. and the MDUMIQE. Further we see that the performance of CondQ and ShiftQ are not sensitive to the choice of the tuning parameter $\gamma$. Any of the value of $\gamma$ between 0.0001 and 0.1, performed well in all the experiments. This makes tuning of the CondQ and ShiftQ algorithms easy.

As expected we observe that for the cases with many quantiles ($K = 19$) or rapid changes ($T=100$), the MDUMIQE performs poorly since the adjusted step size to satisfy the monotone property will be too small to efficiently track the dynamics. The algorithm by Cao et al. performs well for the case with many quantiles ($K = 19$) and slow changes ($T = 1000$). However, with few quantiles, the algorithm performs poorer since the approximation of the cumulative distribution becomes poor. Further, with rapid changes ($T = 100$), the approximation procedure struggles to keep track with the changes in the data stream resulting in poor tracking.
\begin{table}
  \centering
  \begin{tabular}{lcccc}
    \hline
                              & \multicolumn{2}{c}{$K=3$}    &    \multicolumn{2}{c}{$K=19$}     \\\hline
                              &    $T=100$   &      $T=1000$ &         $T=100$  &        $T=1000$\\\hline
ShiftQ, $\gamma = 0.9$        &      0.644   &         0.436 &           0.701  &           0.502\\
ShiftQ, $\gamma = 0.1$        &      0.598   &         0.269 &           0.615  &           0.290\\
ShiftQ, $\gamma = 0.01$       &      0.592   &         0.272 &           0.597  &           0.281\\
ShiftQ, $\gamma = 0.001$      &      0.592   &         0.274 &           0.595  &           0.276\\
ShiftQ, $\gamma = 0.0001$     &      0.597   &         0.284 &           0.597  &           0.282\\
CondQ, $\gamma = 0.9$         &      0.514   &         0.346 &           0.500  &           0.302\\
CondQ, $\gamma = 0.1$         &      0.475   &         0.230 &           0.482  &    \textbf{0.247}\\
CondQ, $\gamma = 0.01$        &\textbf{0.471}&\textbf{0.229} &           0.479  &           0.248\\
CondQ, $\gamma = 0.001$       &      0.472   &\textbf{0.229} &           0.480  &           0.257\\
CondQ, $\gamma = 0.0001$      &      0.479   &         0.242 &   \textbf{0.478} &           0.276\\
MDUMIQE                       &      0.706   &         0.285 &           1.504  &           0.618\\
Cao et al., $c = 10$          &      1.438   &         0.735 &           0.693  &           0.354\\
Cao et al., $c = 1$           &      1.412   &         0.608 &           0.685  &           0.269\\
Cao et al., $c = 0.1$         &      1.435   &         0.623 &           0.692  &           0.311\\ \hline
  \end{tabular}
  \caption{Normal distribution periodic case: Estimation error for the different algorithms under optimal step length. The values in boldface show for each case the algorithm performing the best.}
  \label{tab:5}
\end{table}
\begin{table}
  \centering
  \begin{tabular}{lcccc}
    \hline
                              & \multicolumn{2}{c}{$K=3$}    &    \multicolumn{2}{c}{$K=19$}     \\\hline
                              &    $T=100$   &      $T=1000$ &         $T=100$  &        $T=1000$\\\hline
ShiftQ, $\gamma = 0.9$        &      1.045   &         0.670 &           1.090  &           0.722\\
ShiftQ, $\gamma = 0.1$        &      1.051   &         0.608 &           1.054  &           0.624\\
ShiftQ, $\gamma = 0.01$       &      1.050   &         0.604 &           1.047  &           0.603\\
ShiftQ, $\gamma = 0.001$      &      1.050   &         0.603 &           1.046  &           0.601\\
ShiftQ, $\gamma = 0.0001$     &      1.054   &         0.609 &           1.047  &           0.601\\
CondQ, $\gamma = 0.9$         &      0.685   &         0.472 &           0.698  &           0.448\\
CondQ, $\gamma = 0.1$         &\textbf{0.680}&         0.412 &           0.690  &   \textbf{0.420}\\
CondQ, $\gamma = 0.01$        &\textbf{0.680}& \textbf{0.411}&           0.690  &   \textbf{0.420}\\
CondQ, $\gamma = 0.001$       &      0.681   & \textbf{0.411}&           0.689  &           0.422\\
CondQ, $\gamma = 0.0001$      &      0.686   &         0.420 &   \textbf{0.677} &           0.430\\
MDUMIQE                       &      1.255   &         0.567 &           2.160  &           1.200\\
Cao et al., $c = 10$          &      2.049   &         1.487 &           1.387  &           0.607\\
Cao et al., $c = 1$           &      2.181   &         1.380 &           1.377  &           0.597\\
Cao et al., $c = 0.1$         &      2.237   &         1.395 &           1.379  &           0.607\\ \hline
  \end{tabular}
  \caption{Normal distribution switch case: Estimation error for the different algorithms under optimal step length. The values in boldface show the algorithms resulting in minimum estimation error.}
  \label{tab:6}
\end{table}
\begin{table}
  \centering
  \begin{tabular}{lcccc}
    \hline
                              & \multicolumn{2}{c}{$K=3$}    &    \multicolumn{2}{c}{$K=19$}     \\\hline
                              &    $T=100$   &      $T=1000$ &         $T=100$  &        $T=1000$\\\hline
ShiftQ, $\gamma = 0.9$        &      1.573   &         1.252 &           1.791  &           1.506\\
ShiftQ, $\gamma = 0.1$        &      1.265   &         0.656 &           1.362  &           0.751\\
ShiftQ, $\gamma = 0.01$       &      1.220   &         0.653 &           1.254  &           0.702\\
ShiftQ, $\gamma = 0.001$      &      1.222   &         0.670 &           1.246  &           0.693\\
ShiftQ, $\gamma = 0.0001$     &      1.310   &         0.798 &           1.293  &           0.768\\
CondQ, $\gamma = 0.9$         &      1.350   &         1.026 &           1.237  &           0.870\\
CondQ, $\gamma = 0.1$         &      1.085   & \textbf{0.572}&           1.103  &   \textbf{0.675}\\
CondQ, $\gamma = 0.01$        &\textbf{1.052}&         0.584 &           1.077  &           0.683\\
CondQ, $\gamma = 0.001$       &      1.053   &         0.591 &    \textbf{1.069}&           0.647\\
CondQ, $\gamma = 0.0001$      &      1.128   &         0.692 &           1.187  &           0.680\\
MDUMIQE                       &      1.291   &         0.625 &           1.453  &           0.745\\
Cao et al., $c = 10$          &      1.750   &         1.669 &           1.374  &           0.687\\
Cao et al., $c = 1$           &      1.569   &         1.403 &           1.465  &           0.748\\
Cao et al., $c = 0.1$         &      1.662   &         1.487 &           1.477  &           0.918\\ \hline
  \end{tabular}
  \caption{$\chi^2$ distribution periodic case: Estimation error for the different algorithms under optimal step length. The values in boldface show the algorithms resulting in minimum estimation error.}
  \label{tab:7}
\end{table}
\begin{table}
  \centering
  \begin{tabular}{lcccc}
    \hline
                              & \multicolumn{2}{c}{$K=3$}    &    \multicolumn{2}{c}{$K=19$}     \\\hline
                              &    $T=100$   &      $T=1000$ &         $T=100$  &        $T=1000$\\\hline
ShiftQ, $\gamma = 0.9$        &      1.882   &         1.411 &           2.081  &           1.640\\
ShiftQ, $\gamma = 0.1$        &      1.638   &         0.939 &           1.725  &           1.045\\
ShiftQ, $\gamma = 0.01$       &      1.603   &         0.985 &           1.647  &           1.037\\
ShiftQ, $\gamma = 0.001$      &      1.605   &         0.998 &           1.642  &           1.032\\
ShiftQ, $\gamma = 0.0001$     &      1.676   &         1.093 &           1.678  &           1.083\\
CondQ, $\gamma = 0.9$         &      1.578   &         1.173 &           1.502  &           1.028\\
CondQ, $\gamma = 0.1$         &      1.383   &  \textbf{0.815}&          1.407  &   \textbf{0.905}\\
CondQ, $\gamma = 0.01$        &\textbf{1.361}&         0.857 &           1.389  &           0.938\\
CondQ, $\gamma = 0.001$       &      1.362   &         0.868 &   \textbf{1.386} &           0.917\\
CondQ, $\gamma = 0.0001$      &      1.424   &         0.944 &           1.498  &           0.980\\
MDUMIQE                       &      1.669   &         0.922 &           2.072  &           1.199\\
Cao et al., $c = 10$          &      2.359   &         2.277 &           1.878  &           1.048\\
Cao et al., $c = 1$           &      2.100   &         1.937 &           2.017  &           1.124\\
Cao et al., $c = 0.1$         &      2.196   &         2.009 &           1.977  &           1.293\\ \hline
  \end{tabular}
  \caption{$\chi^2$ distribution switch case: Estimation error for the different algorithms under optimal step length. The values in boldface show the algorithms resulting in minimum estimation error.}
  \label{tab:8}
\end{table}

\subsection{Real-life Data Streams -- Activity Change Detection}

Activity recognition is a highly active field of research where the goal is to use sensors to automatically detect and identify the activities a user is performing. E.g. one could identify if a user (perhaps an obese child) is performing a healthy amount of exercise. We will focus on identifying changes in activities using accelerometer data which is available on almost any smart cell phone today. 

We consider an accelerometer dataset from the Wireless Sensor Data Mining (WISDM) project \cite{kwapisz2011activity}. Accelerations in $x$, $y$ and $z$ directions where observed, with a frequency of 20 observations per second, while users where performing the activities walking, jogging, walking up a stairway and walking down a stairway. A total of 36 users were observed and the dataset contained a total of $989\,875$ observations. 

Current research focuses on supervised approaches where historic and annotated activity observations are used to train an activity classification model \cite{bulling2014tutorial}. For instance, the work reported in \cite{kwapisz2011activity} trained models like decision trees and neural networks. However such an approach is highly sensitive to changes over time like if the user changes to an activity that is not part of the training material, becomes fitter, sick etc. In this example, we rather take an unsupervised approach and the goal is to detect whenever the user changes activity. Since we receive 20 accelerometer observations per second, it is important that the streaming approach is computationally efficient.

Change detection is useful as part of a sequential supervised scheme. Whenever a change is detected, the observations from the last activity can be classified and the supervised classifier retrained. If the supervised learner is sufficiently uncertain about the activity type of the last activity, it may ask the user for feedback in an online manner to gradually improve performance.

\begin{figure}
  \centering
    \includegraphics[width = \textwidth]{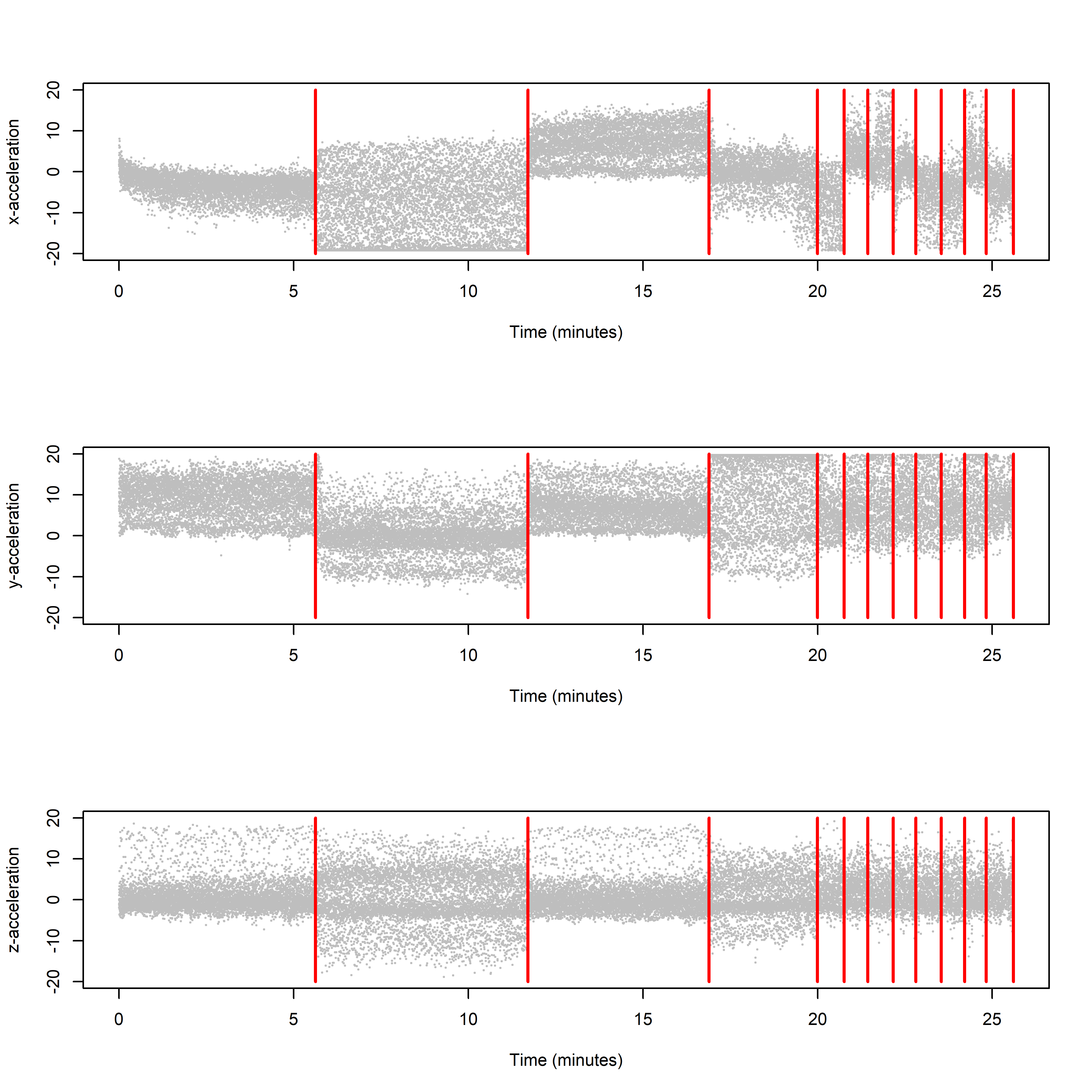}
    \caption{The gray dots show accelerometer observations for an arbitrary users. The red lines show when the user changes activity.}
  \label{fig:22}
\end{figure}

Figure \ref{fig:22} shows in gray $x$, $y$ and $z$ acceleration for an arbitrary user. The red lines show when the user changed activity. Mostly the acceleration distributions are fairly stationary within an activity, but with some gradual and abrupt changes. The users often changed activities as often as every 30 second making this a challenging tracking and change detection problem. 

We suggest the following change detection procedure. Let
$\widehat{Q}_{n,w}(q)$ denote the estimate of the quantile associated with probability $q$ of the accelerometer observations at time $n$ and in dimension $w \in \{x,y,z\}$. 
\begin{enumerate}
\item Use a multiple quantile tracking algorithm to obtain estimates $\widehat{Q}_{n,w}(q_k), k=1,\ldots,K, w = x,y,z$.
\item In each dimension compute the Euclidean distance between the current quantile estimates and the estimates $h$ seconds back in time
  \begin{align*}
    \text{ED}_{n,w} = \sqrt{ \sum_{k=1}^K \left(\widehat{Q}_{n,w}(q_k) - \widehat{Q}_{n-h,w}(q_k) \right)^2 } 
  \end{align*}
\item In each dimension characterize the main distributional properties of $\text{ED}_{n,w}$ by tracking the first two moments using exponentially weighted moving averages
  \begin{align*}
    \hat{\mu}(\text{ED}_{n,w}) &= (1 - \xi) \hat{\mu}(\text{ED}_{n-1,w}) + \xi \text{ED}_{n,w} \\
    \hat{\mu}(\text{ED}_{n,w}^2) &= (1 - \xi) \hat{\mu}(\text{ED}_{n-1,w}^2) + \xi \text{ED}_{n,w}^2 \\
    \hat{\sigma}(\text{ED}_{n,w}) &= \sqrt{\hat{\mu}(\text{ED}_{n,w})^2 - \hat{\mu}(\text{ED}_{n,w}^2)}
  \end{align*}
\item When the user changes activity, we expect ED$_{n,w}$ to rapidly increase in at least one dimension $w$. Thus a new activity is detected when ED$_{n,w}$ is more than $\eta$ standard deviations higher then $\hat{\mu}(\text{ED}_{n,w})$ in at least one dimension, i.e. if
  \begin{align*}
    \max_w \left\{ \frac{\text{ED}_{n,w} - \hat{\mu}(\text{ED}_{n,w})}{\hat{\sigma}(\text{ED}_{n,w})} \right\} \geq \eta
  \end{align*}
\item When a new activity was detected, restart the tracking of the quantile estimates and go back to step 1. 
\end{enumerate}

We compare the quantile tracking approach above with tracking the first two moments of the acceleration distribution in each dimension $w \in \{x,y,z\}$ leading to the following approach:
\begin{enumerate}
\item Let $x_{n,w}$ denote the observed accelerations at time $n$ in dimension $w \in \{x,y,z\}$. Track the mean and standard deviation in each dimension using exponentially weighted moving average
  \begin{align}
    \label{eq:20} \hat{\mu}(X_{n,w}) &= (1 - \nu) \hat{\mu}(X_{n,w}) + \nu x_{n,w}  \\
    \label{eq:21} \hat{\mu}(X^2_{n,w}) &= (1 - \nu) \hat{\mu}(X^2_{n,w}) + \nu x^2_{n,w} \\
    \label{eq:22} \hat{\sigma}(X_{n,w}) &= \sqrt{\hat{\mu}(\text{ED}_{n,w})^2 - \hat{\mu}(\text{ED}_{n,w}^2)}      
  \end{align}
\item In each dimension compute the Mahalanobis distance (MD) between the current estimate of the mean and the estimate $h$ seconds back in time, $\text{MD}_{n,w} = |\hat{\mu}(X_{n,w}) - \hat{\mu}(X_{n-h,w})|/\hat{\sigma}(X_{n,w})$.
\item In each dimension characterize the main distributional properties of $\text{MD}_{n,w}$ by tracking the first two moments using exponentially weighted moving averages
  \begin{align*}
    \hat{\mu}(\text{MD}_{n,w}) &= (1 - \xi) \hat{\mu}(\text{MD}_{n-1,w}) + \xi \text{MD}_{n,w} \\
    \hat{\mu}(\text{MD}_{n,w}^2) &= (1 - \xi) \hat{\mu}(\text{MD}_{n-1,w}^2) + \xi \text{MD}_{n,w}^2 \\
    \hat{\sigma}(\text{MD}_{n,w}) &= \sqrt{\hat{\mu}(\text{MD}_{n,w})^2 - \hat{\mu}(\text{MD}_{n,w}^2)}
  \end{align*}
\item When the user changes activity, we expect MD$_{n,w}$ to rapidly increase in at least one dimension $w$. Thus a new activity is detected when MD$_{n,w}$ where more than $\eta$ standard deviations higher then $\hat{\mu}(\text{MD}_{n,w})$ in at least one dimension, i.e. if
  \begin{align*}
    \max_w \left\{ \frac{\text{MD}_{n,w} - \hat{\mu}(\text{MD}_{n,w})}{\hat{\sigma}(\text{MD}_{n,w})} \right\} \geq \eta
  \end{align*}
\item When a new activity is detected, restart the tracking of the quantile estimates and go back to step 1. 
\end{enumerate}

A disadvantage with the MD approach above is that it only can detect changes in the first two moments of the acceleration distributions, while the quantile tracking approach can detect \textit{any} changes in the distributions. 

We measured the performance of the approaches for a wide range of values for the tuning parameters. To properly characterize the acceleration distributions, we tracked a total of $K = 9$ quantiles, namely quantiles associated with the probabilities $0.1, 0.2, \ldots, 0.9$.
Given the scale of the observations small values of $\lambda$, $\beta$ and $\nu$ are reasonable and we used 0.001, 0.005 and 0.01. Further we used 0.005, 0.01 and 0.05 for $\xi$, 5, 10 and 15 seconds for $h$ we used and 10, 15 and 25 for $\eta$. Finally, following the recommendations in \cite{hammer2018a}, we used $\rho/\lambda = 0.01$. We ran the change detection approaches for the whole dataset for all the combinations of the tuning parameters resulting in a total of 162 and 81 experiments for the quantile and MD approaches, respectively. 

To measure detection performance we used the well-known measures precision, recall and the F1 score \cite {sokolova2009systematic}. If the approach detects more than one change between two true changes, we characterize the first change as a correct detection and the others as false detections. Then define precision, recall and F1 score 
\begin{align*}
  \text{Precision} &= \frac{ \text{No. of correct detections} }{ \text{No. of detections} } \\[2mm]
  \text{Recall} &= \frac{ \text{No. of correct detections} }{ \text{No. true changes} } \\[2mm]
  \text{F1 score} &= \frac{ 2\,\cdot\text{Precision}\,\cdot\, \text{Recall}  }{\text{Precision} + \text{Recall}} 
\end{align*}
where the F1 score is the harmonic mean of precision and recall.

Tables \ref{tab:1} to \ref{tab:4} show the top ten results with respect to the F1 score for the different approaches. We see that the CondQ and ShiftQ outperform  MD. Further we see that CondQ detects true changes more rapidly than ShiftQ (last column). Using CondQ for change detection seems to be the best alternative. 
\begin{table}
  \centering
  \begin{tabular}{cccc|cccc}
    \hline
    $\nu$ & $\xi$ & $h$ & $\eta$ & Precision & Recall & F1 score & Det. delay \\ \hline
    0.010  & 0.010  & 5   &    25  &     0.633  &  0.639  & \textbf{0.636}  &          \textbf{15.469}\\       
    0.005  & 0.005  & 10  &    25  &     0.707  &  0.578  & \textbf{0.636}  &          \textbf{25.072}\\       
    0.010  & 0.005  & 10  &    25  &     0.686  &  0.592  & \textbf{0.636}  &          \textbf{22.637}\\       
    0.005  & 0.010  & 5   &    25  &     0.617  &  0.642  & \textbf{0.629}  &          \textbf{16.739}\\       
    0.010  & 0.010  & 15  &    25  &     0.668  &  0.592  & \textbf{0.628}  &          \textbf{30.128}\\       
    0.010  & 0.010  & 10  &    25  &     0.616  &  0.639  & \textbf{0.627}  &          \textbf{16.795}\\       
    0.010  & 0.005  & 5   &    25  &     0.690  &  0.566  & \textbf{0.622}  &          \textbf{19.326}\\       
    0.005  & 0.010  & 15  &    25  &     0.630  &  0.595  & \textbf{0.612}  &          \textbf{36.359}\\       
    0.005  & 0.010  & 10  &    25  &     0.581  &  0.642  & \textbf{0.610}  &          \textbf{24.175}\\       
    0.005  & 0.005  & 5   &    25  &     0.666  &  0.558  & \textbf{0.607}  &          \textbf{15.683}\\ \hline
  \end{tabular}
  \caption{Change detection example. Results for the MD approach. Detection delay (last column) is given in seconds.}
  \label{tab:1}
\end{table}
\begin{table}
  \centering
  \begin{tabular}{ccccc|cccc}
    \hline
$\lambda$ & $\gamma$ & $\xi$ &  $h$ & $\eta$ & Precision & Recall & F1 score & Det. delay \\ \hline
0.001   &    0.2  &  0.05  & 10  &    15  &   0.694  &  0.662  & \textbf{0.678}  &          \textbf{18.615}\\       
0.001   &    0.1  &  0.05  & 10  &    25  &   0.781  &  0.587  & \textbf{0.670}  &          \textbf{16.229}\\       
0.001   &    0.1  &  0.05  & 15  &    25  &   0.798  &  0.572  & \textbf{0.667}  &          \textbf{13.993}\\       
0.001   &    0.2  &  0.01  & 10  &    10  &   0.795  &  0.572  & \textbf{0.666}  &          \textbf{18.441}\\       
0.001   &    0.1  &  0.01  & 5   &    10  &   0.727  &  0.607  & \textbf{0.661}  &          \textbf{26.102}\\       
0.001   &    0.1  &  0.05  & 10  &    15  &   0.592  &  0.746  & \textbf{0.660}  &          \textbf{16.727}\\       
0.001   &    0.2  &  0.05  & 15  &    15  &   0.714  &  0.613  & \textbf{0.659}  &          \textbf{26.152}\\       
0.001   &    0.1  &  0.01  & 10  &    10  &   0.690  &  0.604  & \textbf{0.644}  &          \textbf{18.833}\\       
0.001   &    0.1  &  0.05  & 5   &    15  &   0.621  &  0.662  & \textbf{0.641}  &          \textbf{21.371}\\       
0.001   &    0.1  &  0.01  & 15  &    10  &   0.700  &  0.561  & \textbf{0.623}  &          \textbf{23.274}\\ \hline
  \end{tabular}
  \caption{Change detection example. Results using ShiftQ to track quantiles. Detection delay (last column) is given in seconds.}
  \label{tab:3}
\end{table}
\begin{table}
  \centering
  \begin{tabular}{ccccc|cccc}
    \hline 
$\lambda$ & $\gamma$ & $\xi$ &  $h$ & $\eta$ & Precision & Recall & F1 score & Det. delay \\ \hline
0.010  &     0.1  &  0.05  & 10  &    25  &      0.785  &  0.613  & \textbf{0.688}  &          \textbf{15.457}\\       
0.010  &     0.2  &  0.05  & 10  &    15  &      0.705  &  0.662  & \textbf{0.683}  &          \textbf{13.400}\\       
0.010  &     0.1  &  0.01  & 10  &    10  &      0.723  &  0.633  & \textbf{0.675}  &          \textbf{19.437}\\       
0.010  &     0.1  &  0.05  & 15  &    25  &      0.786  &  0.584  & \textbf{0.670}  &          \textbf{18.470}\\         
0.005  &     0.1  &  0.05  & 10  &    25  &      0.792  &  0.572  & \textbf{0.664}  &          \textbf{17.855}\\       
0.010  &     0.2  &  0.05  & 15  &    15  &      0.711  &  0.618  & \textbf{0.662}  &          \textbf{19.957}\\       
0.005  &     0.1  &  0.05  & 15  &    25  &      0.779  &  0.569  & \textbf{0.658}  &          \textbf{22.672}\\       
0.005  &     0.1  &  0.01  & 10  &    10  &      0.720  &  0.595  & \textbf{0.652}  &          \textbf{20.829}\\       
0.005  &     0.2  &  0.05  & 15  &    15  &      0.723  &  0.590  & \textbf{0.650}  &          \textbf{22.142}\\       
0.005  &     0.1  &  0.05  & 15  &    15  &      0.624  &  0.668  & \textbf{0.645}  &          \textbf{27.182}\\ \hline
  \end{tabular}
  \caption{Change detection example. Results using CondQ to track quantiles. Detection delay (last column) is given in seconds.}
  \label{tab:4}
\end{table}

\section{Closing Remark}
\label{sec:closrem}

Incremental quantile estimators document state-of-the performance to track quantiles of dynamically varying data streams. They are however not able to compute joint estimates of multiple quantiles and even the monotone property of quantiles may get violated. In this paper we present a new procedure that is able use incremental quantile estimators to obtain joint estimates of multiple quantiles. The fundamental idea is to first track a central quantile and track other quantiles relative to the central quantile. The joint properties are preserved by using properties of shifted and conditional quantiles.

We apply the procedure to obtain the ShiftQ and CondQ algorithms which are extensions of the DUMIQE and QEWA algorithms, respectively. The experiments show that the ShiftQ and CondQ algorithms outperform state-of-the-art multiple quantile tracking algorithms in both synthetic and real-life data streams. Further, CondQ outperforms ShiftQ which is as expected since the QEWA documents better performance than DUMIQE \cite{hammer2018a}. 

A direction for future research is to apply this general concept of conditional quantiles to also extend other incremental quantile estimators.

\clearpage

\bibliographystyle{plain}
\bibliography{bibl}

\clearpage

\appendix

\section{Synthetic Experiments - Complete Results}
\label{app:synt}

\begin{figure}
  \centering
  \begin{tabular}{cc}
   \includegraphics[width = 0.5\textwidth]{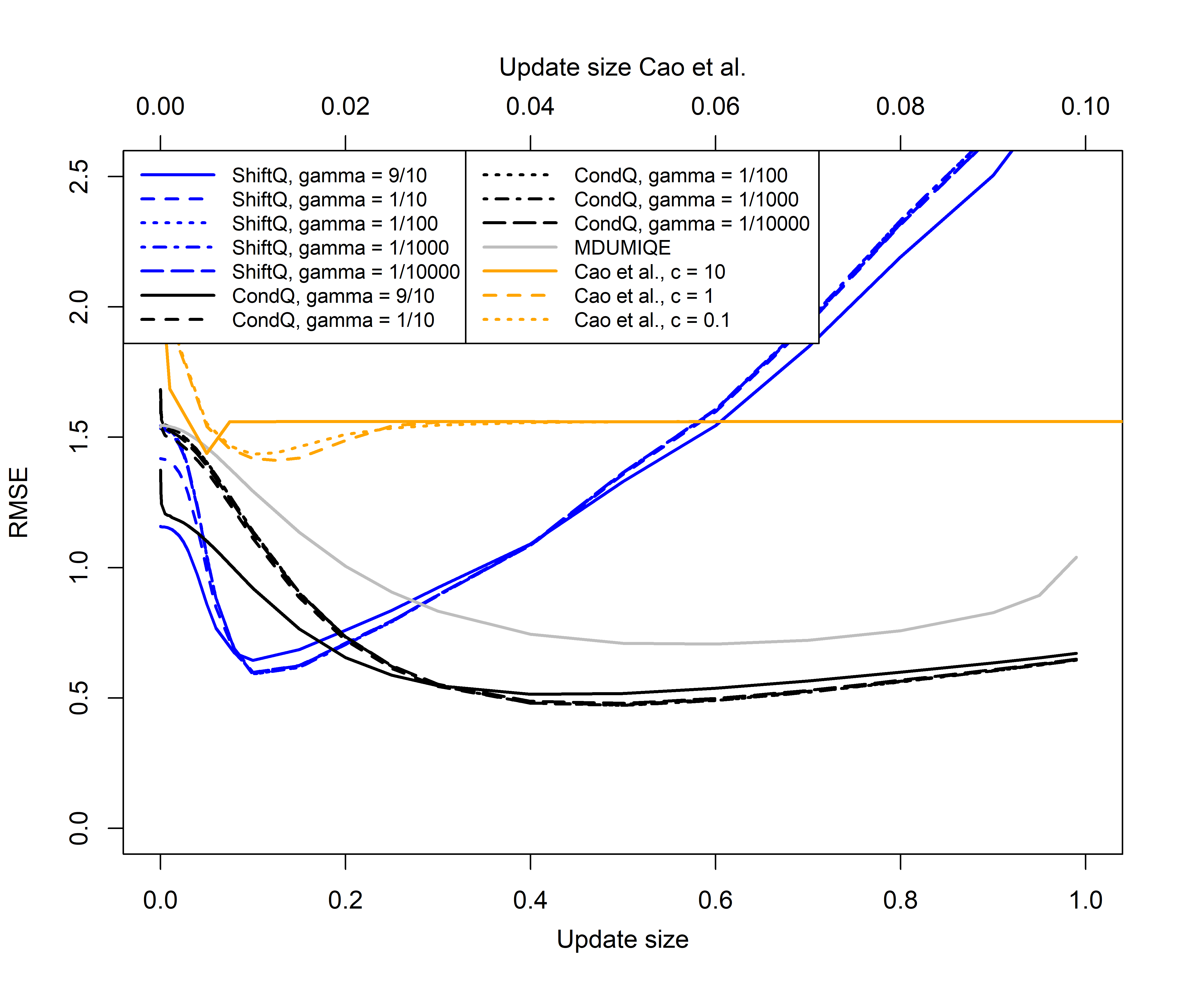} & \includegraphics[width = 0.5\textwidth]{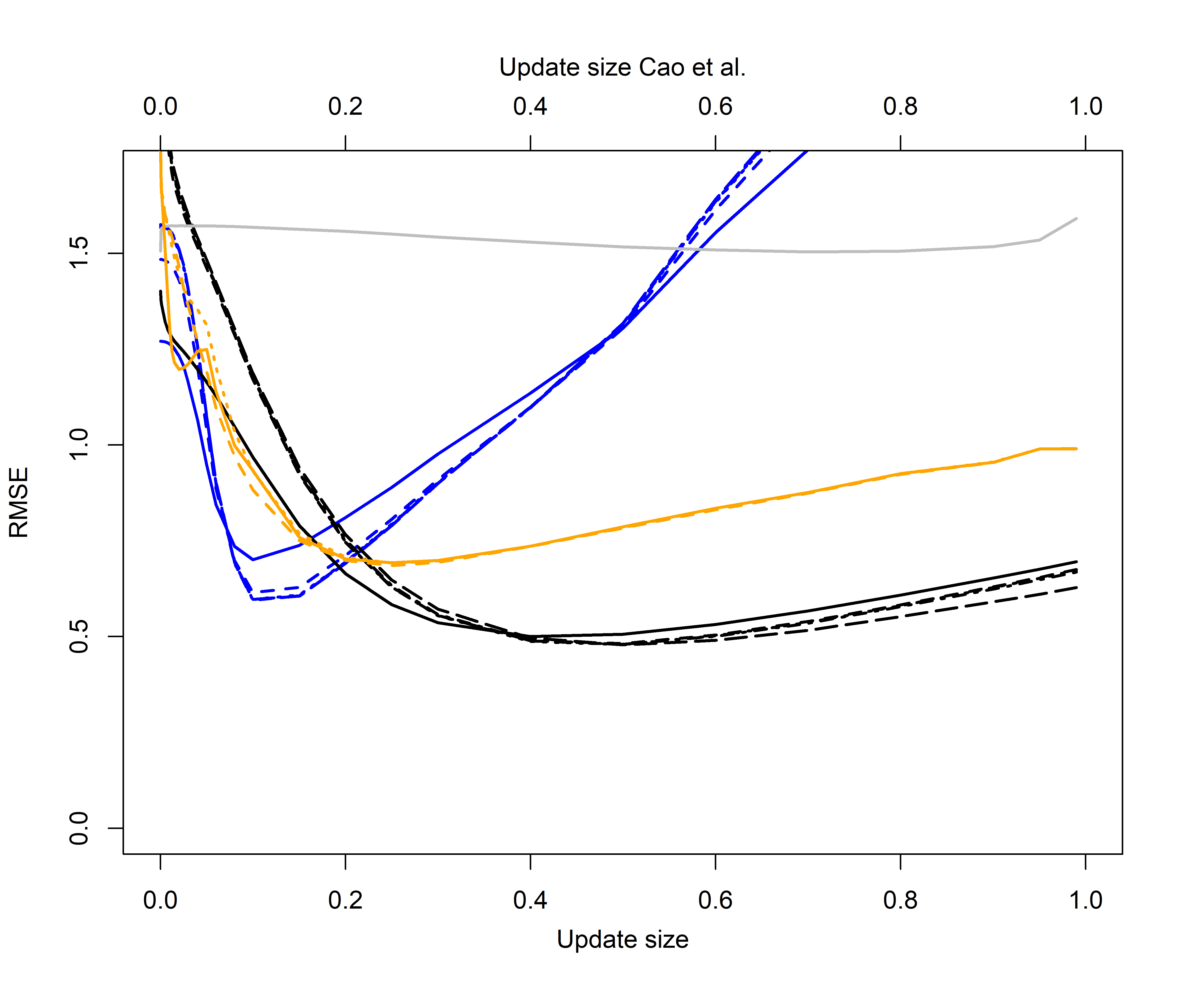} \\
   \includegraphics[width = 0.5\textwidth]{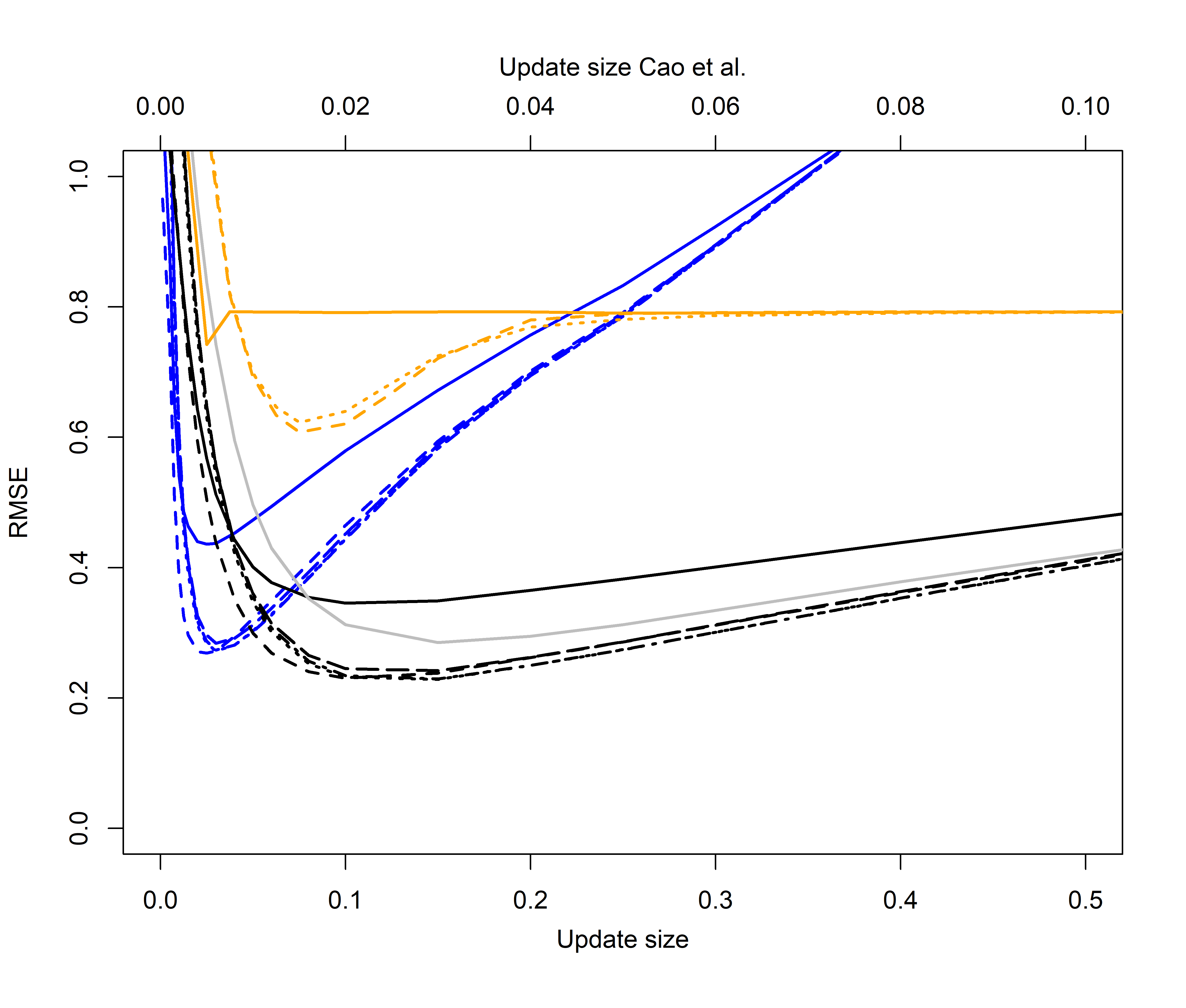} & \includegraphics[width = 0.5\textwidth]{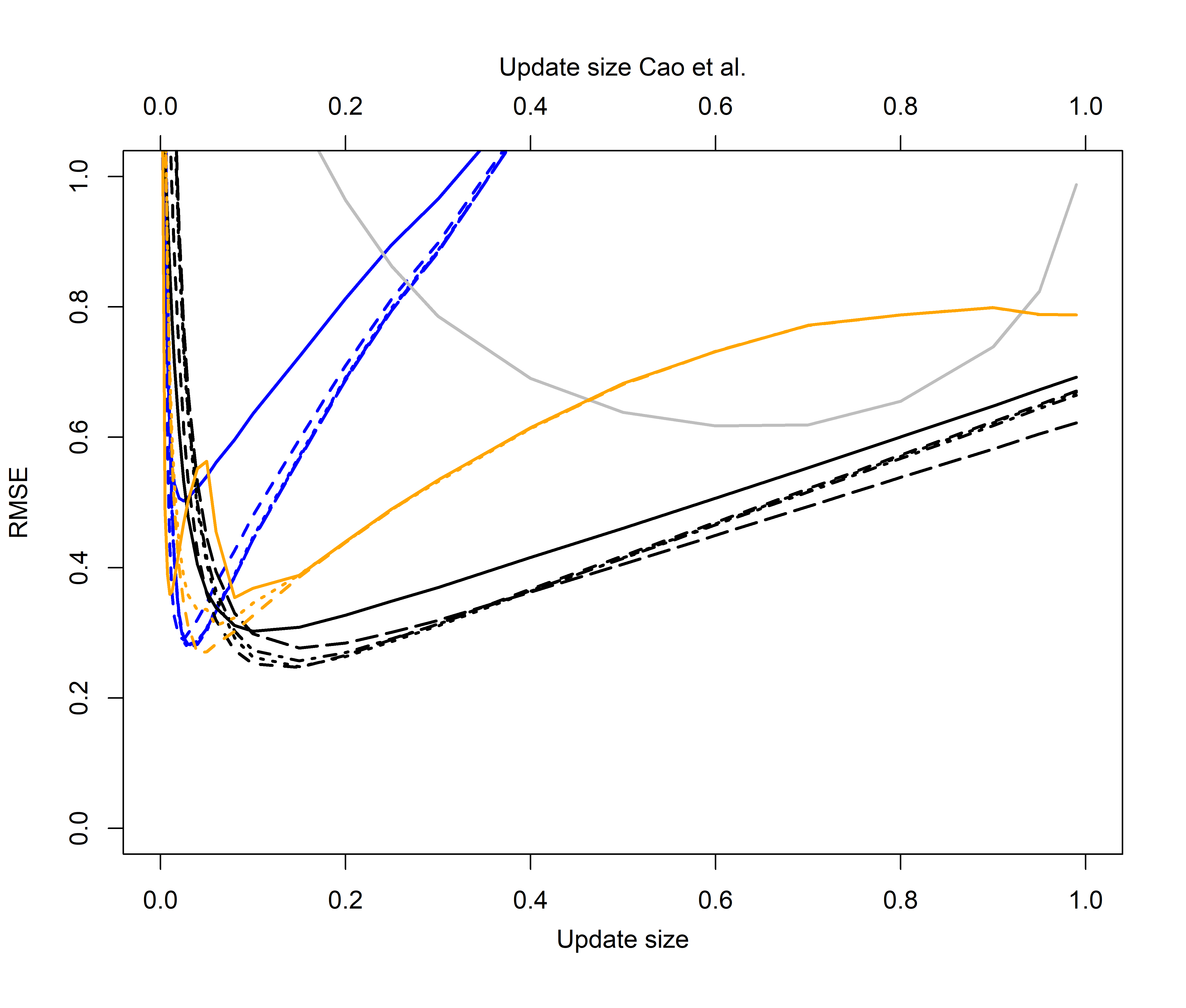} \\
  \end{tabular}
  \caption{Normal distribution periodic case: The left and right columns show results for $K= 3$ and $K = 19$, respectively. The top and bottom rows show results for periods $T = 100$ and $1000$, respectively.}
  \label{fig:2}
\end{figure}

\begin{figure}
  \centering
  \begin{tabular}{cc}
   \includegraphics[width = 0.5\textwidth]{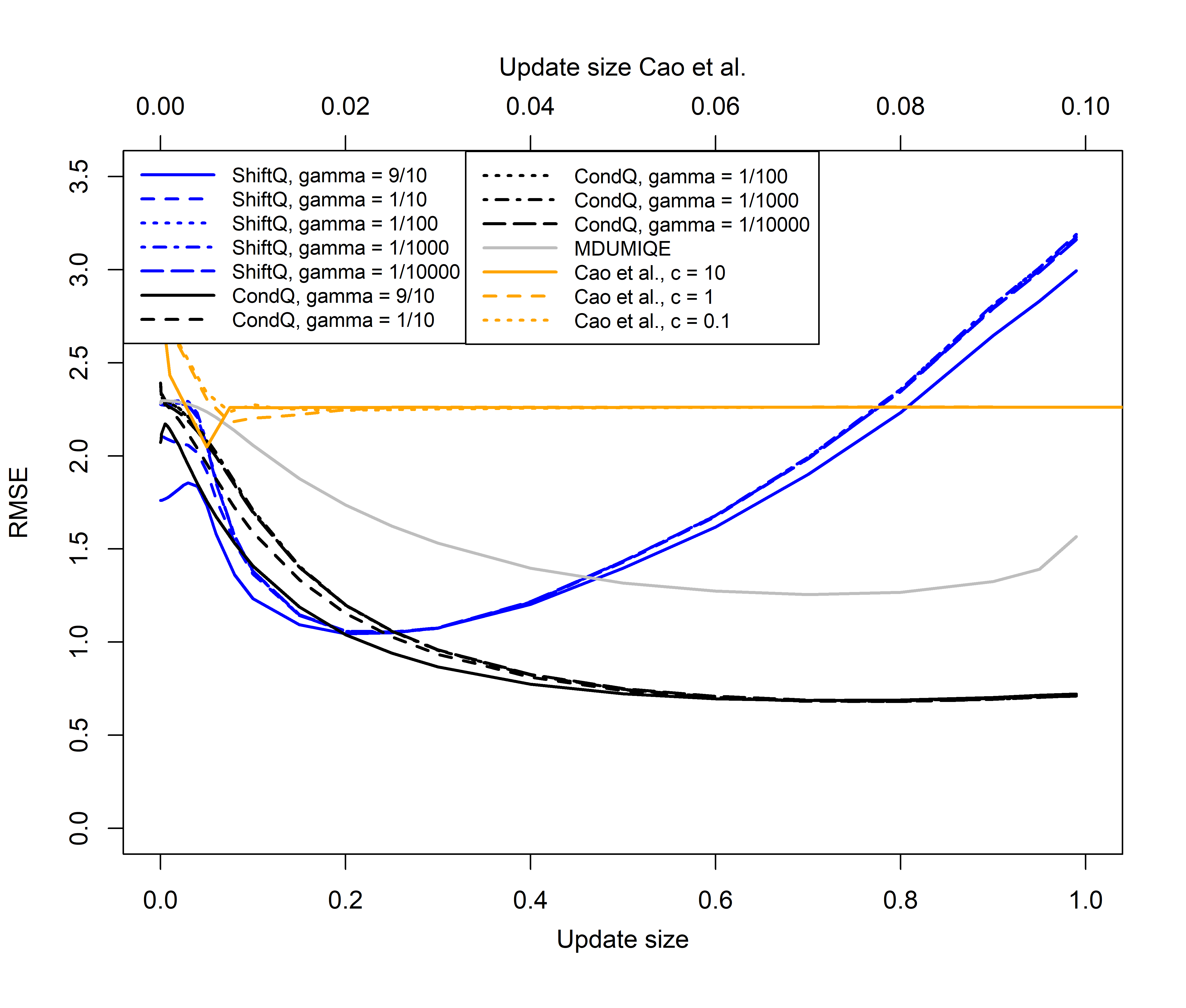} & \includegraphics[width = 0.5\textwidth]{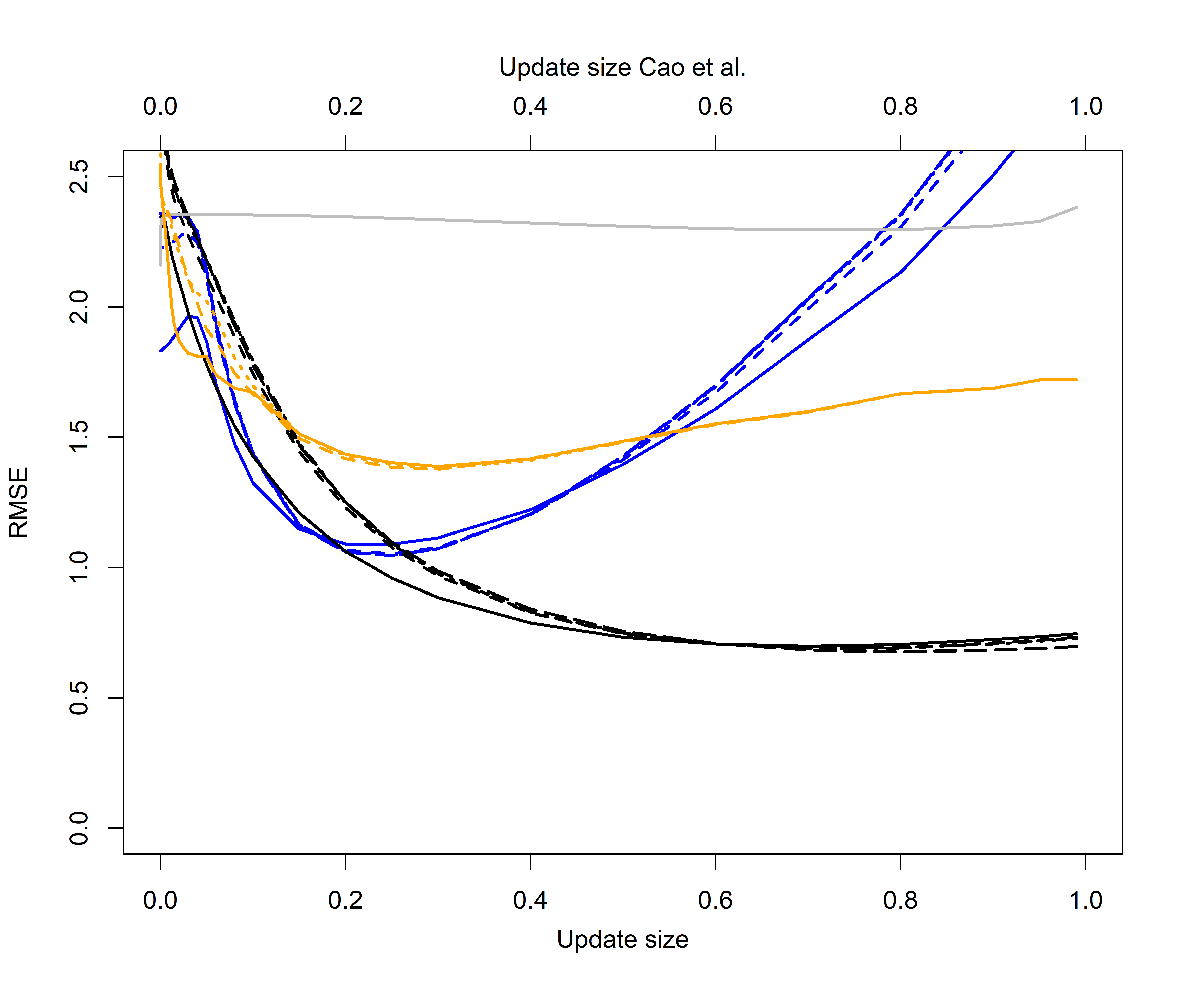} \\
   \includegraphics[width = 0.5\textwidth]{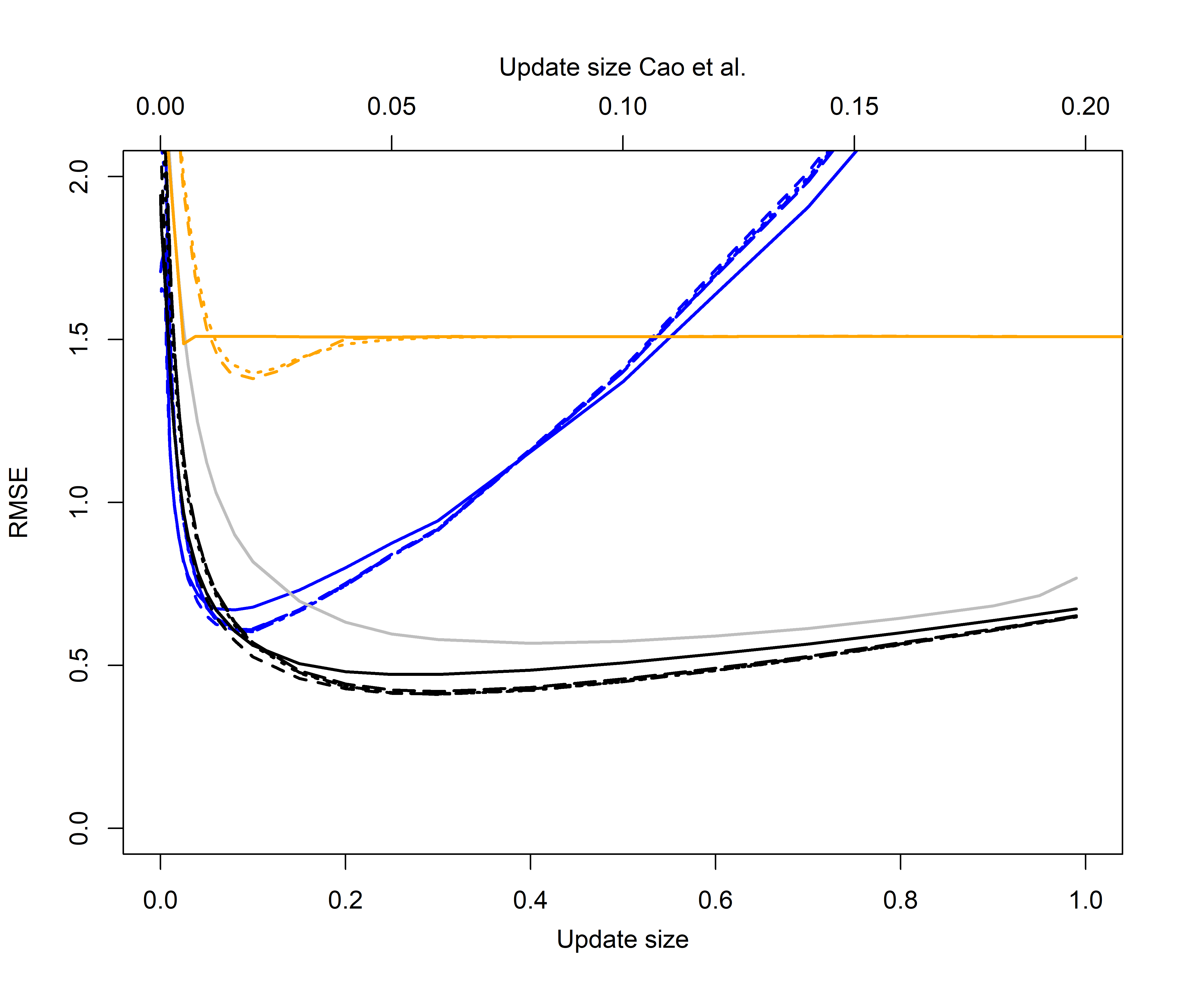} & \includegraphics[width = 0.5\textwidth]{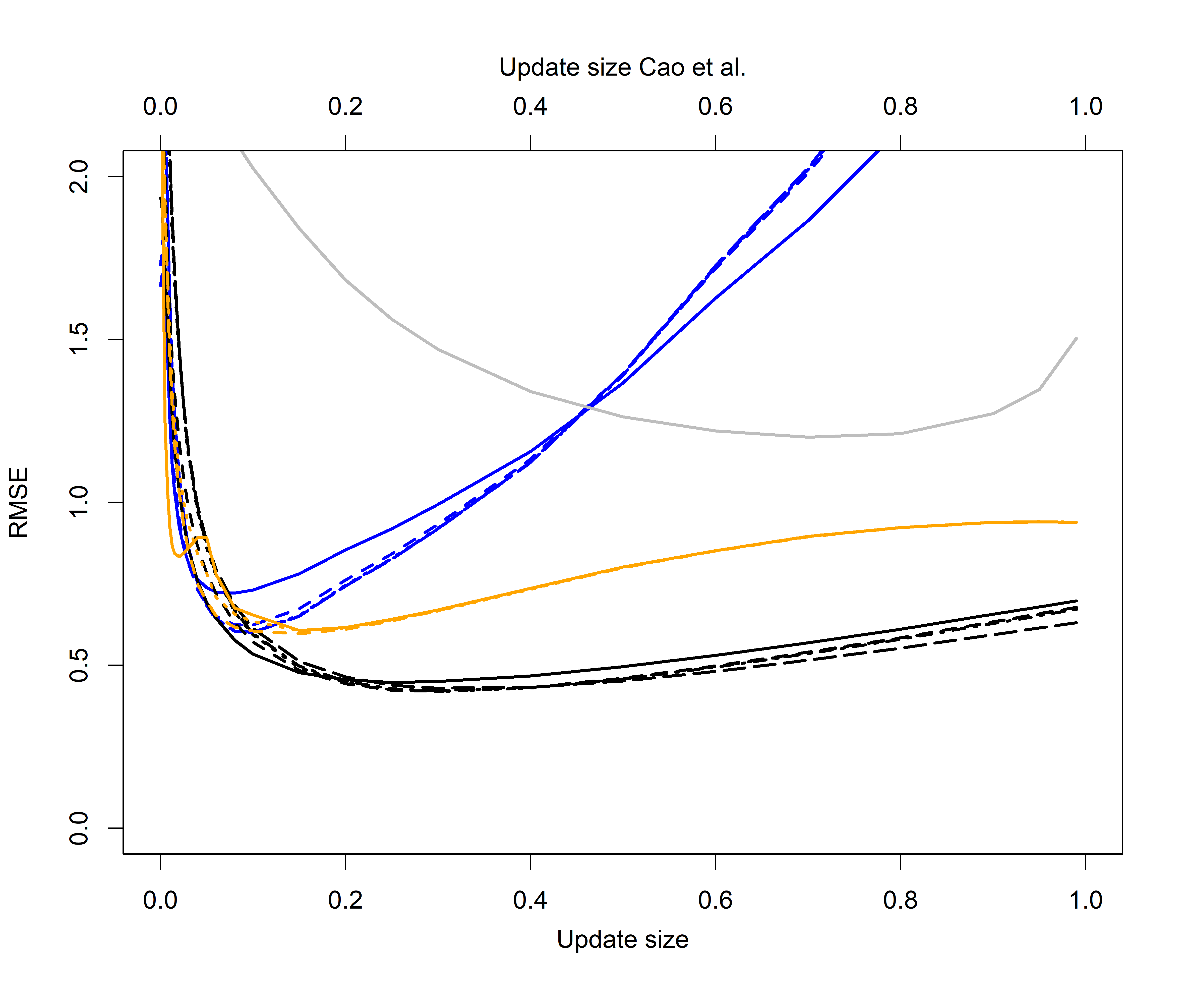} \\
  \end{tabular}
  \caption{Normal distribution switch case: The left and right columns show results for $K= 3$ and $K = 19$, respectively. The top and bottom rows show results for periods $T = 100$ and $1000$, respectively.}
  \label{fig:3}
\end{figure}

\begin{figure}
  \centering
  \begin{tabular}{cc}
   \includegraphics[width = 0.5\textwidth]{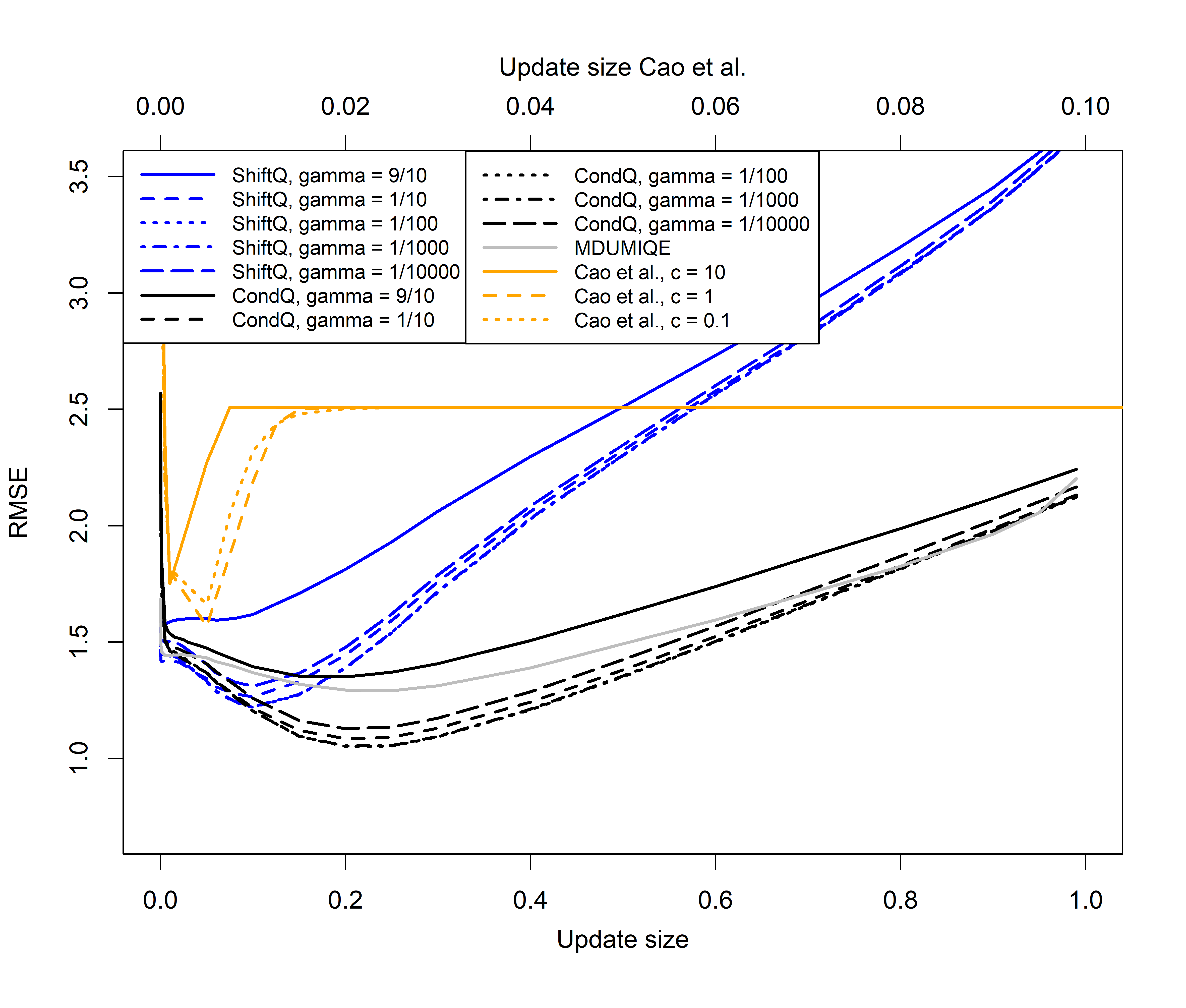} & \includegraphics[width = 0.5\textwidth]{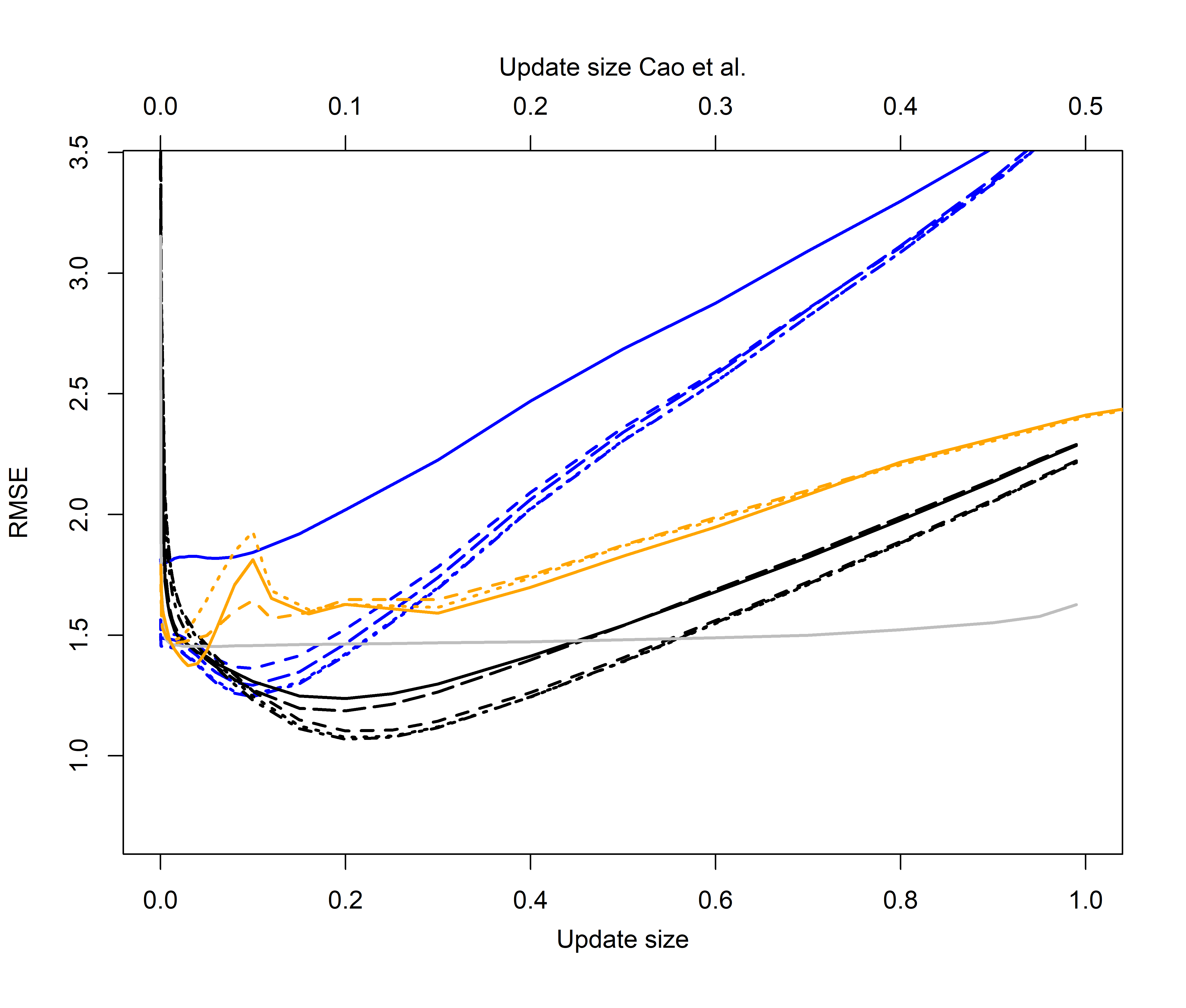} \\
   \includegraphics[width = 0.5\textwidth]{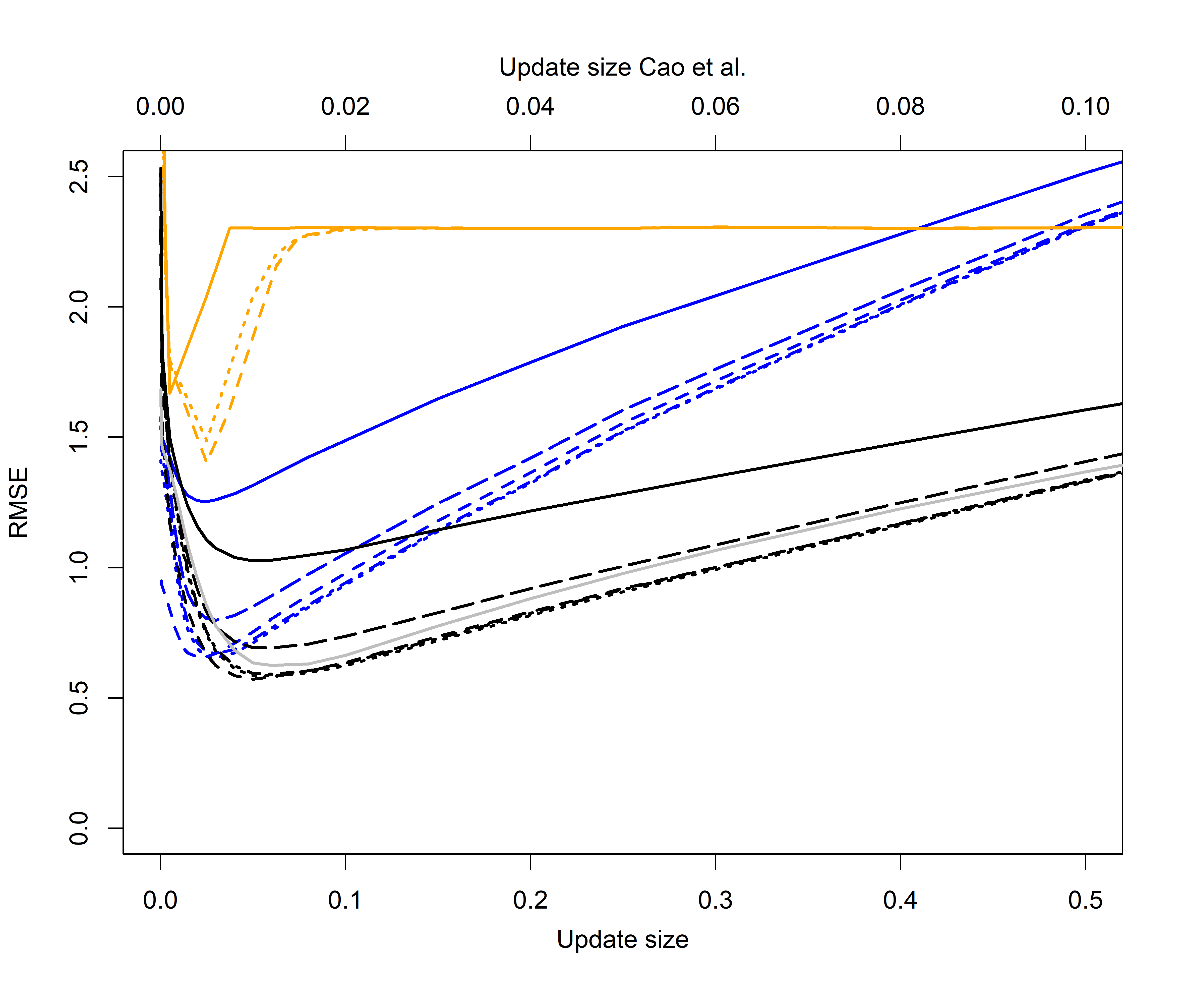} & \includegraphics[width = 0.5\textwidth]{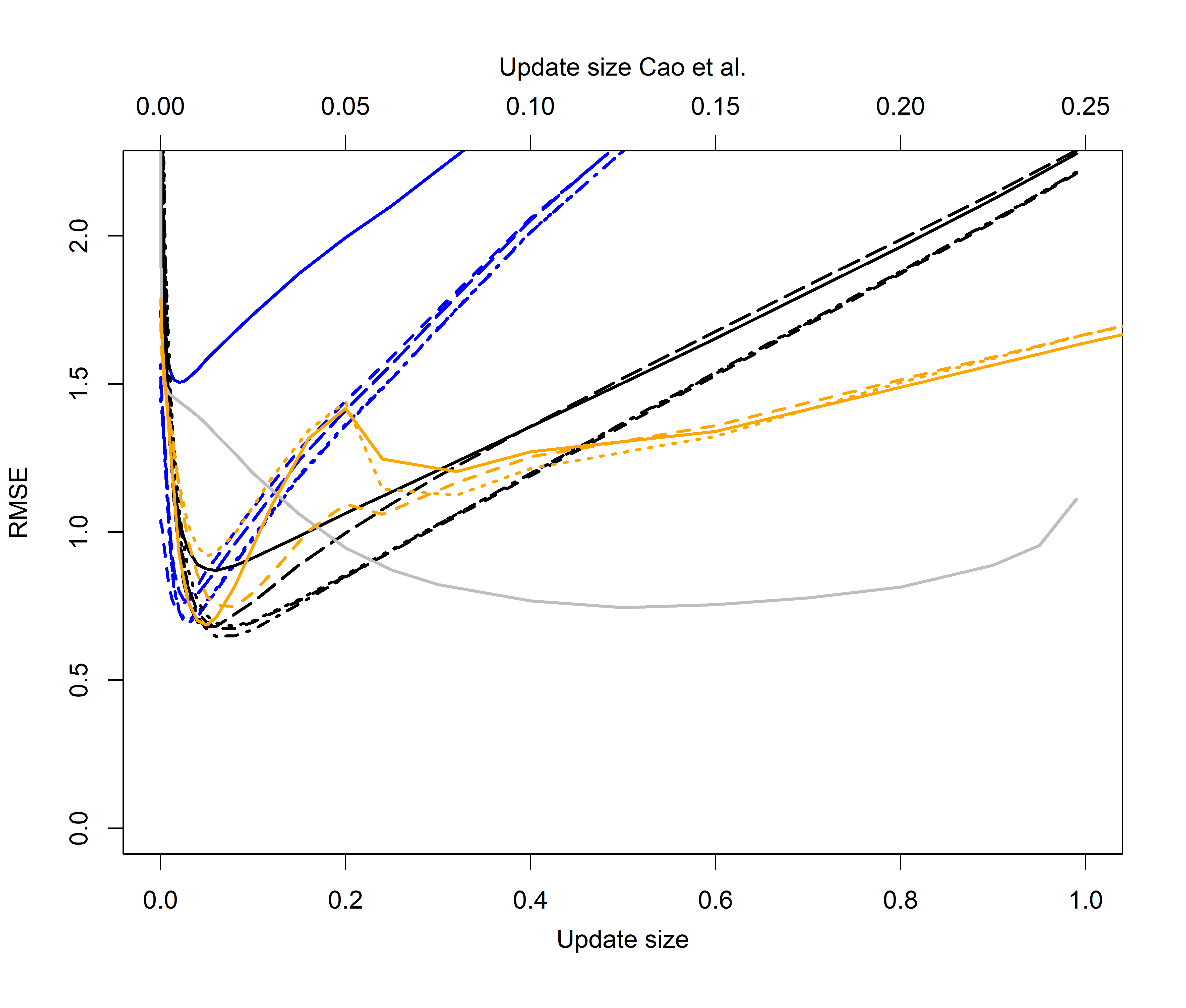} \\
  \end{tabular}
  \caption{$\chi^2$ distribution periodic case: The left and right columns show results for $K= 3$ and $K = 19$, respectively. The top and bottom rows show results for periods $T = 100$ and $1000$, respectively.}
  \label{fig:4}
\end{figure}

\begin{figure}
  \centering
  \begin{tabular}{cc}
   \includegraphics[width = 0.5\textwidth]{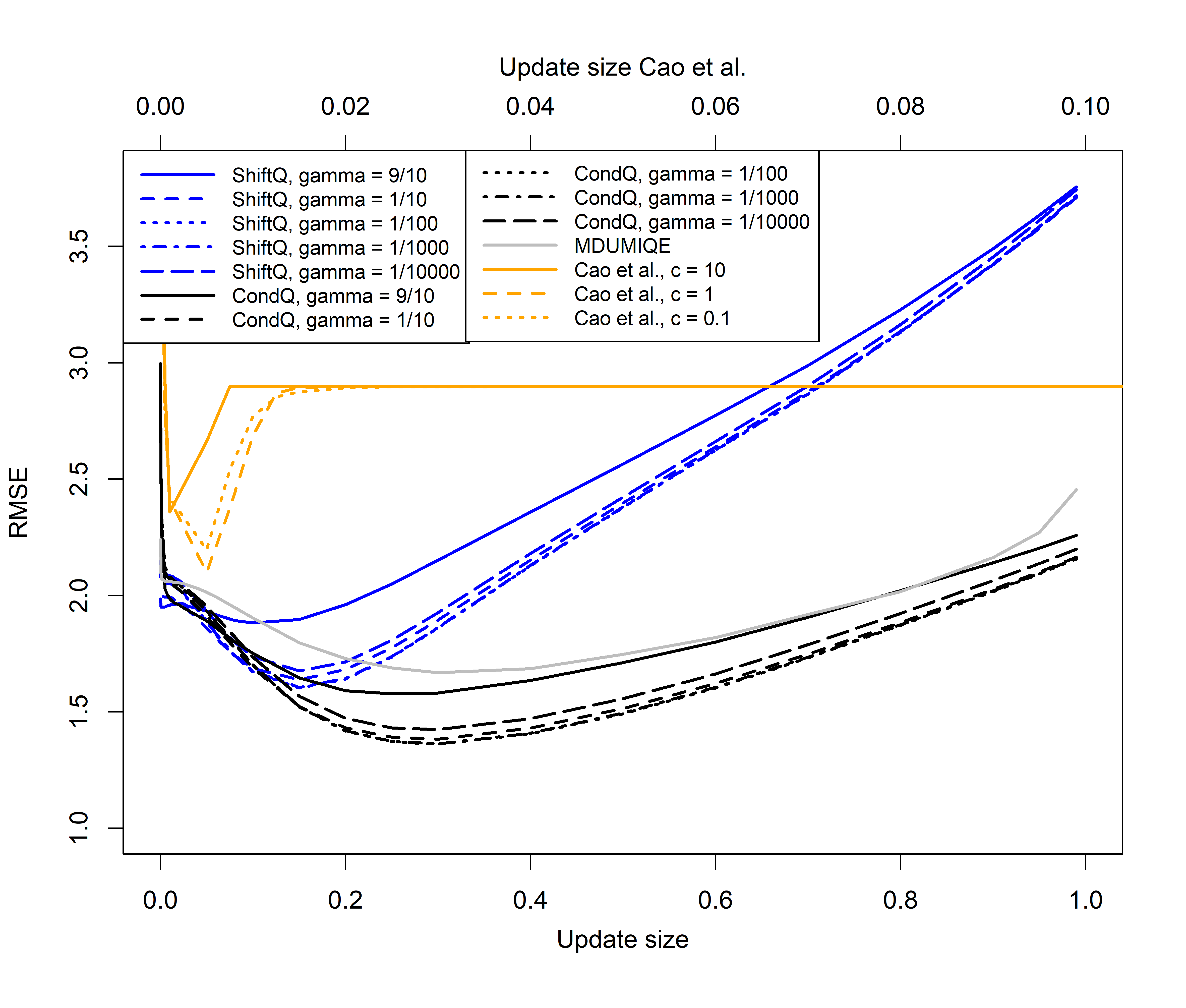} & \includegraphics[width = 0.5\textwidth]{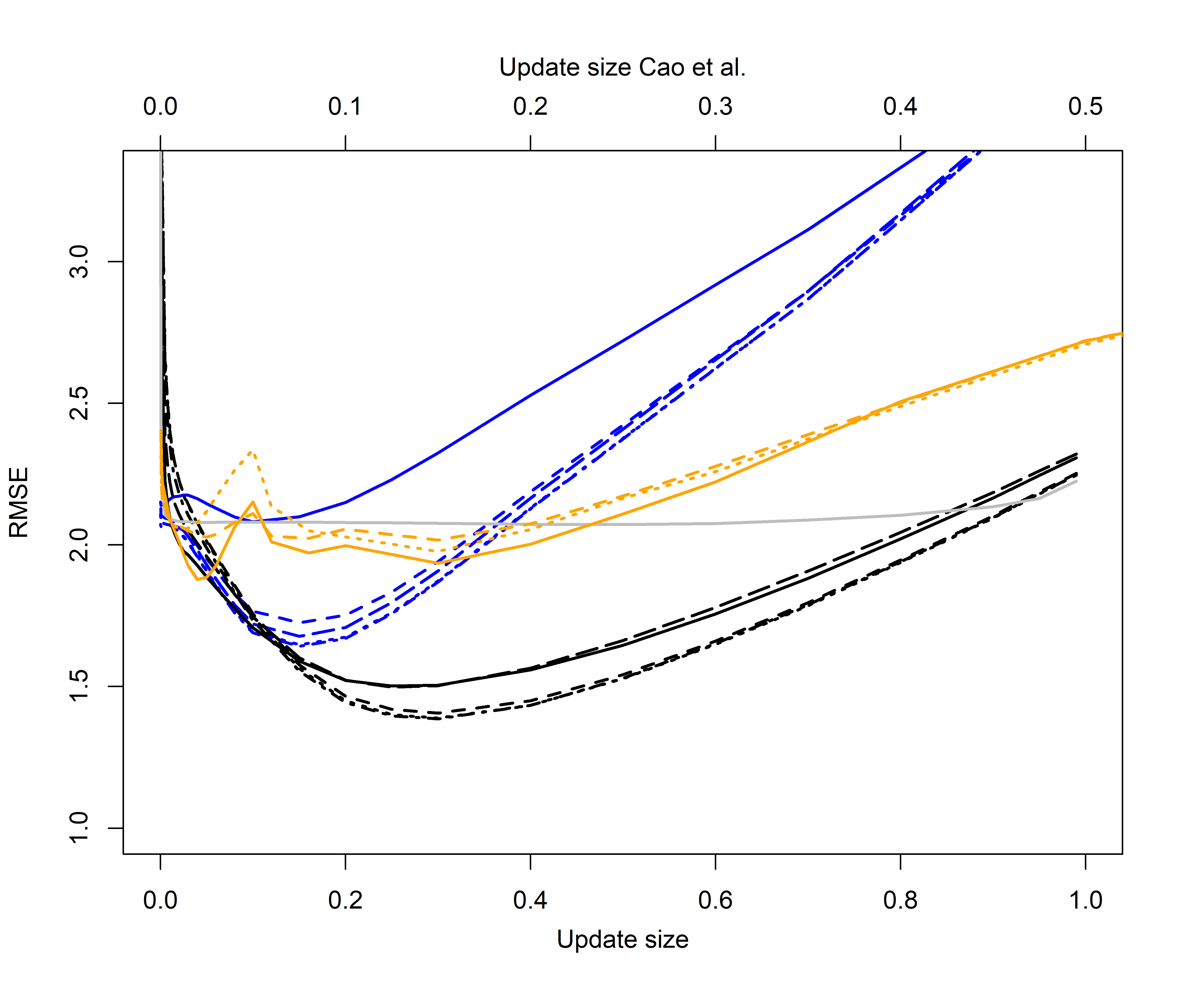} \\
   \includegraphics[width = 0.5\textwidth]{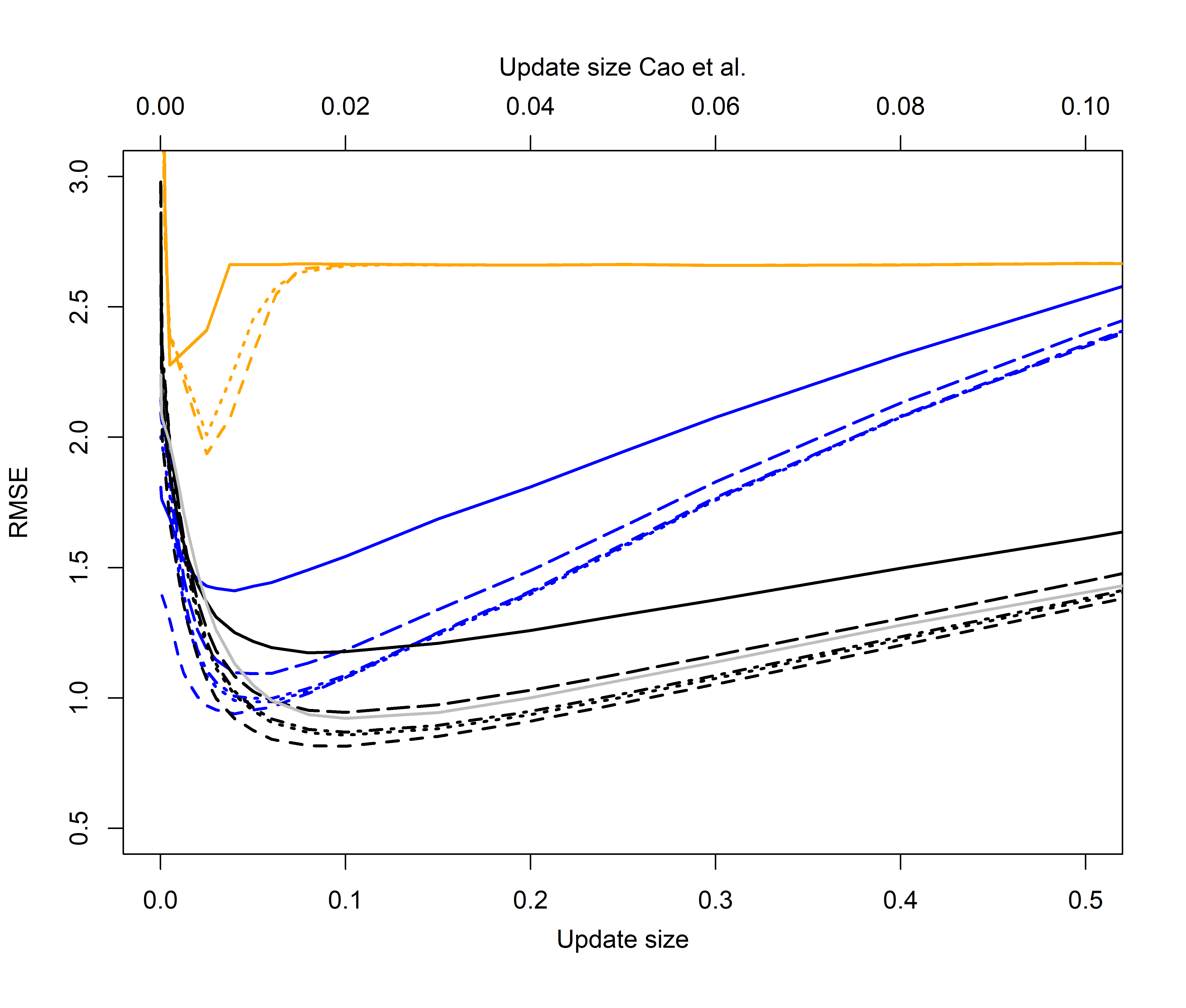} & \includegraphics[width = 0.5\textwidth]{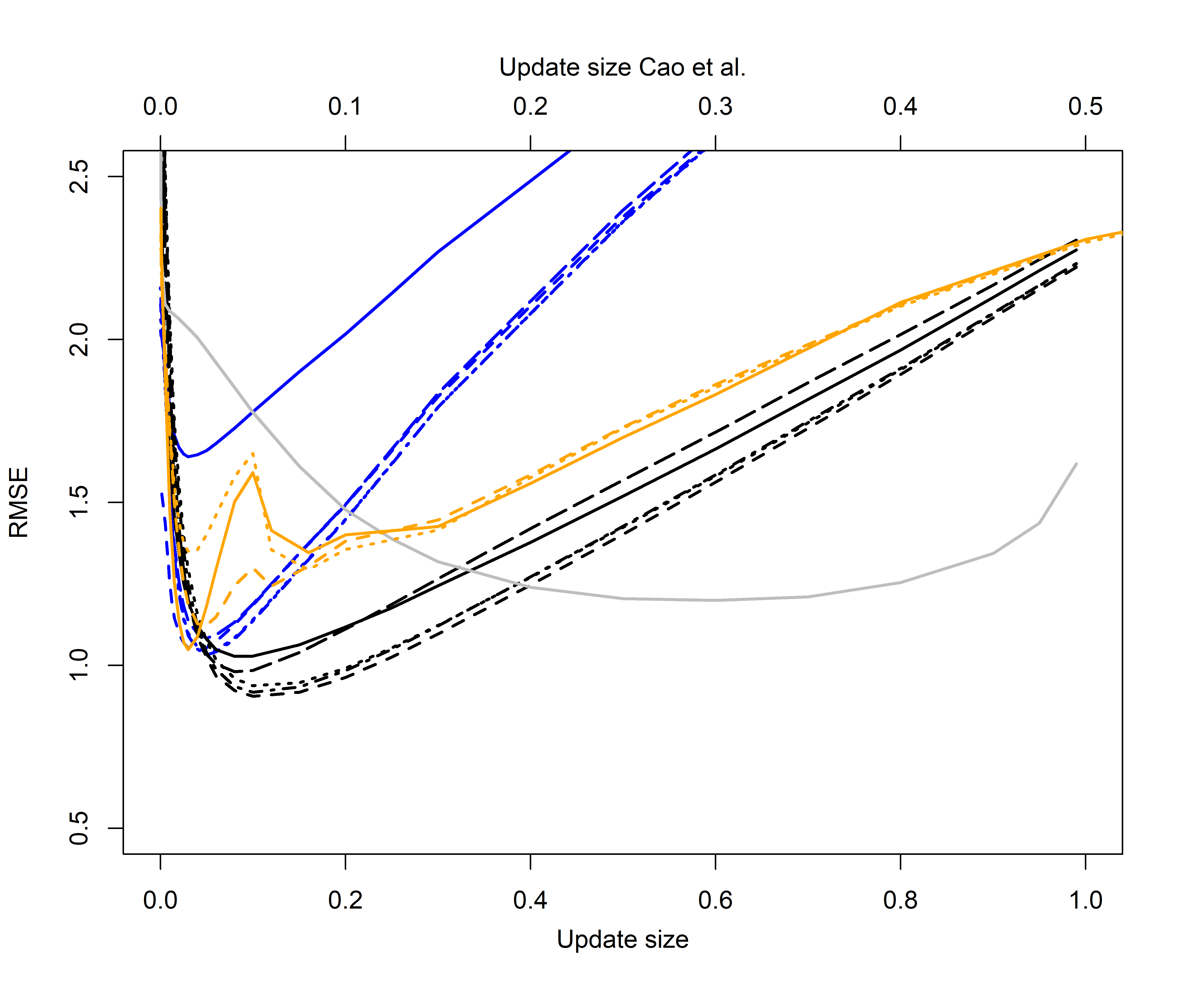} \\
  \end{tabular}
  \caption{$\chi^2$ distribution switch case: The left and right columns show results for $K= 3$ and $K = 19$, respectively. The top and bottom rows show results for periods $T = 100$ and $1000$, respectively.}
  \label{fig:5}
\end{figure}

Figures \ref{fig:2} to \ref{fig:5} show the complete error curves for the synthetic experiments in Section \ref{sec:exp}.

\end{document}